\documentclass[preprint,3p]{elsarticle}

\journal{}
\usepackage{graphicx}
\usepackage{dcolumn}
\usepackage{bm}
\usepackage{bigints}
\usepackage[mathlines]{lineno}
\linenumbers\relax 

\usepackage[utf8]{inputenc}
\usepackage[T1]{fontenc}
\usepackage{amsmath,amssymb}
\usepackage{etoolbox}
\usepackage{bbold}
\usepackage{color}
\usepackage{pgfplots}
\pgfplotsset{compat=newest}
\usepackage{tikz}
\usetikzlibrary{patterns}

\newcommand{\opt}{\mathrm{opt}}

\makeatletter
\def\thickhline{%
  \noalign{\ifnum0=`}\fi\hrule \@height \thickarrayrulewidth \futurelet
   \reserved@a\@xthickhline}
\def\@xthickhline{\ifx\reserved@a\thickhline
               \vskip\doublerulesep
               \vskip-\thickarrayrulewidth
             \fi
      \ifnum0=`{\fi}}
\makeatother

\newlength{\thickarrayrulewidth}
\journal{Journal of Sound and Vibration}

\begin{document}
\nolinenumbers

\begin{frontmatter}

\title{Optimal damping of vibrating systems: dependence on initial conditions}
\author[1]{K. Lelas\corref{cor1}}
\ead{klelas@ttf.unizg.hr}

\author[2]{I. Nakić}
\ead{nakic@math.hr}

\affiliation[1]{
organization={Faculty of Textile Technology, University of Zagreb},
country={Croatia}
}
\affiliation[2]{
organization={Department of Mathematics, Faculty of Science, University of Zagreb},
country={Croatia}
}
\cortext[cor1]{Corresponding author}


\begin{abstract}
Common criteria used for measuring performance of vibrating systems have one thing in common: they do not depend on initial conditions of the system. In some cases it is assumed that the system has zero initial conditions, or some kind of averaging is used to get rid of initial conditions. The aim of this paper is to initiate rigorous study of the dependence of vibrating systems on initial conditions in the setting of optimal damping problems. We show that, based on the type of initial conditions, especially on the ratio of potential and kinetic energy of the initial conditions, the vibrating system will have quite different behavior and correspondingly the optimal damping coefficients will be quite different. More precisely, for single degree of freedom systems and the initial conditions with mostly potential energy, the optimal damping coefficient will be in the under-damped regime, while in the case of the predominant kinetic energy the optimal damping coefficient will be in the over-damped regime. In fact, in the case of pure kinetic initial energy, the optimal damping coefficient is $+\infty$! Qualitatively, we found the same behavior in multi degree of freedom systems with mass proportional damping. We also introduce a new method for determining the optimal damping of vibrating systems, which takes into account the peculiarities of initial conditions and the fact that, although in theory these systems asymptotically approach equilibrium and never reach it exactly, in nature and in experiments they effectively reach equilibrium in some finite time.
\end{abstract}


\end{frontmatter}

\section{Introduction}

If we have an multi-degree of freedom (MDOF) linear vibrating system, i.e. a system of coupled damped oscillators, how to determine damping coefficients that ensure optimal evanescence of free vibrations? 
In the literature one finds several different criteria, typically based on frequency domain analysis of the system, although there are also approaches based on time domain analysis \cite{rojas2019optimal}. The tools used for designing the criteria include modal analysis \cite{fu2001modal},  transfer functions \cite{takewaki1997optimal}, $H_2$ and $H_\infty$ norms coming from systems theory \cite{CHEUNG200929, ALUJEVIC20144073} and spectral techniques \cite{LANCASTER2005891}. A general overview of the optimization tools for structures analysis can be found in e.g.\ \cite{haftka2012elements}.
Another optimization criterion used is to take as optimal the damping coefficients that minimize the (zero to infinity) time integral of the energy of the system, averaged over all possible initial conditions corresponding to the same initial energy \cite{Ves90}. 
This criterion was investigated widely, mostly by mathematicians in the last two decades, more details can be found, e.g., in references \cite{Ves90, cox2004lyapunov, NakicPhd, truhar2009efficient, veselic2011damped}. 

However, what is common to all these criteria is that they implicitly or explicitly ignore the dependence of the dynamics of the system on the initial conditions. 
Sometimes this is suitable, e.g. for systems with continuous excitation, but in some cases it make sense to study the free vibrations of the system with non-zero initial conditions. A prominent example where this is the case is the vibration control of buildings subjected to earthquake excitation \cite{ventura1992influence, wang2007vibration}. 
Indeed, depending on the initial conditions, MDOF systems can exhibit oscillatory or non-oscillatory response \cite{morzfeld2013characterization}, so it is clear that initial conditions can play an important role in the overall dynamics of the system.

Implicitly, dependence of the behavior of system on initial conditions has been investigated in the context of time-optimal vibrations reduction \cite{dhanda2016vibration} and transient response \cite{meirelles2005transient} in terms of computationally efficient methods for the calculation of the system response. Our aim with this paper is to start more systematic investigation of the role of initial conditions in the study of linear vibrating systems. 
Specifically, the dependence of the energy integral on the initial conditions has not been investigated, as far as we are aware, and therefore it is not clear how much information about the behavior of vibrating systems is lost by taking the average over all initial conditions or by assuming zero initial conditions and it is not clear how well the optimal damping obtained in this way works for a specific choice of initial conditions, e.g. for an experiment with initial conditions such that the initial energy consists only of potential energy, etc. We have chosen to study the particular criterion
of minimizing time integral of the energy as in this case it is straightforward to modify it to take into account the initial conditions: instead of averaging over all possible initial conditions, we study the dependence of the time integral of the energy of the system on initial conditions. Specifically, for criteria based on frequency domain approach, which are designed for forced vibrations, it is not clear how to take into account the non-zero initial conditions in a systematic way. 

\medskip

We will explore this dependence by considering free vibrations of single degree of freedom (SDOF), two-degree of freedom (2-DOF) and MDOF vibrating systems with mass proportional damping (MPD). In particular, for a SDOF, averaging over all initial conditions gives the critical damping as optimal \cite{Ves90,NakicPhd}, and we show, by considering the minimization of the energy integral without averaging over initial conditions, that damping coefficients approximately $30\%$ less than critical to infinite are obtained as optimal, depending on the initial conditions. We systematize all our results with respect to the relationship between initial potential and initial kinetic energy, e.g., for initial conditions with initial potential energy grater than initial kinetic energy the optimal damping coefficient is in the under-damped regime, while for initial conditions with initial kinetic energy grater than initial potential energy we find the optimal damping deep in the over-damped regime. We also consider the minimization of the energy integral averaged over a subset of initial conditions and obtain a significant dependence of the optimal damping coefficient on the selected subset. Qualitatively, we find the same behavior in 2-DOF and MDOF systems as well.

Furthermore, we show that the minimization of the energy integral for certain types of initial conditions does not give a satisfactory optimal damping coefficient. Specifically, for SDOF systems, the obtained optimal damping coefficient does not distinguish between two initial states with the same magnitude of initial displacement and initial velocity, but which differ in the relative sign of initial displacement and initial velocity. These initial conditions differ significantly in the rate of energy dissipation as a function of the damping coefficient, i.e.\ it is not realistic for one damping coefficient to be optimal for both of these initial conditions. The same is true for each individual mode of MDOF systems with respect to the signs of initial displacements and velocities, expressed via modal coordinates. Another disadvantage of this criterion is that, for initial conditions with purely kinetic initial energy, it gives an infinite optimal damping coefficient, which is not practical for experiments. Also, the energy integral is calculated over the entire time, due to the fact that these systems asymptotically approach equilibrium and never reach it exactly, but in nature and experiments they effectively reach equilibrium in some finite time.

We introduce a new method for determining the optimal damping of MDOF systems, which practically solves the aforementioned problems and gives optimal damping coefficients that take into account the peculiarities of each initial condition and the fact that these systems effectively reach equilibrium in some finite time. We take that the system has effectively reached equilibrium when its energy drops to some small fraction of the initial energy, e.g., to the energy resolution of the measuring device with which we observe the system. Our method is based on the determination of the damping coefficient for which the energy of the system drops to that desired energy level the fastest. 

In this paper we focus on mass proportional damping so that we could analytically perform a modal analysis and present ideas in the simplest possible way, but, as we briefly comment at the end of the paper, everything we have done can be done in a similar fashion analytically for the case of Rayleigh damping \cite{katsikadelis2020dynamic} as well as for tuned mass damper \cite{gutierrez2013tuned,ZUO2004893}. Also, it is possible to carry out this kind of analysis numerically for systems with damping that does not allow analytical treatment. This will be the subject of our further research.

The rest of the paper is organized as follows: Section 2 is devoted to SDOF systems, in particular minimization of the energy integral and optimal damping is studied for the chosen set of initial conditions. In Section 3 we analyze 2-DOF systems with MPD. MDOF systems with MPD are the subject of Section 4. In Section 5 we propose a new optimization criterion and analyze its properties. Section 6 summarizes important findings of the paper.
 

\section{SDOF systems} 
\label{1Dsection}

Free vibrations of a SDOF linear vibrating system can be described by the equation 
\begin{equation}
\ddot x(t)+2\gamma\dot x(t)+\omega_0^2x(t)=0,\, x(0)=x_0, \, \dot{x}(0)=v_0,
\label{DHOeq}
\end{equation}
where $x(t)$ denotes the displacement from the equilibrium position (set to $x=0$) as a function of time, the dots denote time derivatives, $\gamma>0$ is the damping coefficient, $\omega_0$ stands for the undamped oscillator angular frequency (sometimes called the natural frequency of the oscillator)  and $(x_0, v_0)$ encode the initial conditions \cite{grigoriu2021linear, Berkeley}. The physical units of the displacement $x(t)$ depend on the system being considered. For example, for a mass on a spring (or a pendulum) in viscous fluid, when it is usually called \emph{elongation}, it is measured in $[m]$, while for an RLC circuit it could either be voltage, or current, or charge. In contrast, the units of $\gamma$ and $\omega_0$ are $[s^{-1}]$ for all systems described with the SDOF model. The form of the solution to the equation \eqref{DHOeq} depends on the relationship between $\gamma$ and $\omega_0$, producing three possible regimes \cite{grigoriu2021linear, Berkeley}: under-damped ($\gamma<\omega_0$), critically damped ($\gamma=\omega_0$) and over-damped ($\gamma>\omega_0$) regime. 

Here we would like to point out that, although it is natural to classify the solution of SDOF into three regimes depending on the value of $\gamma$, we can actually take one form of the solution as a unique solution valid for all values of $\gamma>0$, $\gamma \ne \omega_0$, 
%
\begin{equation}
x(t)=e^{-\gamma t}\left(x_0\cos(\omega t)+\frac{v_0+\gamma x_0}{\omega}\sin(\omega t)\right)\,,
\label{xud}
\end{equation}
where $\omega=\sqrt{\omega_0^2-\gamma^2}$ is the (complex valued) damped angular frequency. 
%
In order to describe the critically damped regime, 
one can take
the limit $\gamma\rightarrow\omega_0$ of the solution \eqref{xud} to obtain the general solution of the critically damped regime 
\begin{equation}
x_c(t)=e^{-\omega_0 t}\left(x_0+(v_0+\omega_0 x_0)t\right)\,.
\label{xcd}
\end{equation}
Therefore, in order to calculate the energy and the time integral of the energy, we do not need to perform separate calculations for all three regimes, but a single calculation using the displacement given by \eqref{xud} and the velocity given by
\begin{equation}
\dot{x}(t)=e^{-\gamma t}\left(v_0\cos(\omega t)-\frac{\gamma v_0+\omega_0^2 x_0}{\omega}\sin(\omega t)\right)\,.
\label{vud}
\end{equation}

For simplicity, in this section we will refer to the quantity
\begin{equation}
E(t)=\dot{x}(t)^2+\omega_0^2x(t)^2
\label{Et}
\end{equation}
as the \emph{energy} of the system, and to the quantities $E_K(t)=\dot{x}(t)^2$ and $E_P(t)=\omega_0^2x(t)^2$ as the \emph{kinetic energy} and \emph{potential energy} of the system respectively. The connection of the quantity \eqref{Et} to the usual expressions for the energy is straightforward, e.g., for a mass $m$ on a spring in viscous fluid 
\begin{equation}
\mathcal{E}(t)=\frac{m}{2}E(t)\,,
\label{Ephy}
\end{equation}
and similarly for other systems described with the SDOF model. Using \eqref{xud} and \eqref{vud} in \eqref{Et}, we obtain    
\begin{equation}
E(t)=e^{-2\gamma t}\left( E_0\cos^2(\omega t)+\gamma\left(\omega_0^2x_0^2-v_0^2\right)\frac{\sin(2\omega t)}{\omega}+\left(E_0(\omega_0^2+\gamma^2)+4\omega_0^2\gamma x_0v_0\right)\frac{\sin^2(\omega t)}{\omega^2}\right)
\label{Et1}
\end{equation}
for the energy of the system, where $E_0=v_0^2+\omega_0^2x_0^2$ is the initial energy given to the system at $t=0$. Accordingly, $E_{0K}=v_0^2$ is the initial kinetic energy and $E_{0P}=\omega_0^2x_0^2$ is the initial potential energy. Expression \eqref{Et1} is valid for both under-damped and over-damped regimes, and to obtain the energy of the critically damped regime we take the $\gamma\rightarrow\omega_0$ limit of the energy \eqref{Et1}, and obtain
\begin{equation}
E_c(t)=e^{-2\omega_0 t}\left( E_0+2\omega_0\left(\omega_0^2x_0^2-v_0^2\right)t+2\omega_0^2\left(E_0+2\omega_0x_0v_0\right)t^2\right)\,.
\label{Etc}
\end{equation}

\subsection{Minimization of the energy integral and optimal damping in dependence of initial conditions}
\label{1Dinitialcond}

We consider the SDOF system with initially energy $E_0$. All possible initial conditions that give this energy can be expressed in polar coordinates with constant radius $r=\sqrt{E_0}$ and angle $\theta=\arctan\left(\frac{v_0}{\omega_0x_0}\right)$, i.e. we have  
\begin{equation}
\begin{split}
\omega_0x_0=r\cos\theta\\
v_0=r\sin\theta\,.
\label{polar}
\end{split}
\end{equation}
In Fig.\ \ref{fig:skica} we sketch the circle given by \eqref{polar}, i.e. given by all possible initial conditions with the same energy $E_0$. For clarity of the exposition, here we comment on a few characteristic points of the circle presented in Fig.\ \ref{fig:skica}: 
\begin{itemize}
\item Initial conditions $\omega_0x_0=\pm\sqrt{E_0}$ and $v_0=0$, i.e. with purely potential initial energy (and zero initial kinetic energy), correspond to two points on the circle with $\theta=\lbrace{0, \pi\rbrace}$.
\item Initial conditions $\omega_0x_0=\pm\sqrt{E_0/2}$ and $v_0=\pm\sqrt{E_0/2}$, i.e. with initial potential energy equal to initial kinetic energy, correspond to four points on the circle with $\theta=\lbrace\pi/4, 3\pi/4, 5\pi/4, 7\pi/4\rbrace$.
\item Initial conditions $\omega_0x_0=0$ and $v_0=\pm\sqrt{E_0}$, i.e. with purely kinetic initial energy (and zero initial potential energy), correspond to two points on the circle with $\theta=\lbrace{\pi/2, 3\pi/2\rbrace}$.
\end{itemize}   
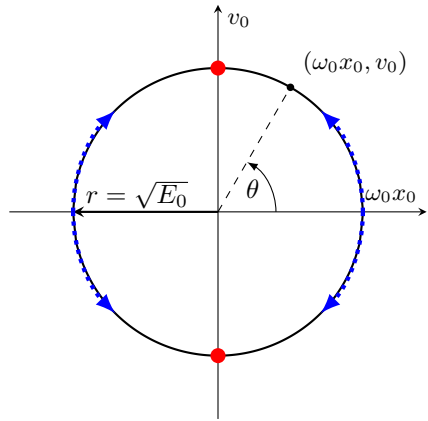
\begin{figure}[h!t!]
\begin{center}
\begin{tikzpicture}
    \begin{axis}[
    axis y line=middle, 
    axis x line=middle,
        width=0.43\textwidth,
        height=0.43\textwidth,
        xmin = -1.8,
        xmax = 1.8,
        ymin = -1.8,
        ymax = 1.8,
        xtick=\empty, 
        ytick=\empty,
        every tick label/.append style={font=\small},
        xlabel = {\small $\omega_0x_0$},
        ylabel = {\small $v_0$}, 
        legend entries ={}
    ]
    \addplot [smooth, domain=-1.25:1.25,samples=100,thick, black](\x,{sqrt(1.25^2-\x^2)});
    \addplot [smooth, domain=-1.25:1.25,samples=100,thick, black](\x,{-sqrt(1.25^2-\x^2)});
    \addplot[dashed, color=black] coordinates {(0,0) (1.25*0.5,1.25*0.866)};
    \draw[-latex] (0:0.5) arc (0:60:0.5);
    \node[] at (axis cs: 0.3,0.2) {$\theta$};
    \draw[->,>=stealth, thick] (axis cs: 0,0) -- (axis cs: -1.25,0);
    \node[] at (axis cs: -0.7,0.15) {$r=\sqrt{E_0}$};
    \node[fill,circle,inner sep=1pt,label={above right:\small $(\omega_0x_0,v_0)$}] at (1.25*0.5,1.25*0.866){};
    \draw[-latex, dotted, blue,  ultra thick] (0:1.25) arc (0:45:1.25);
    \draw[-latex, dotted, blue,  ultra thick] (0:1.25) arc (0:-45:1.25);
    \draw[-latex, dotted, blue,  ultra thick] (0:-1.25) arc (0:45:-1.25);
    \draw[-latex, dotted, blue,  ultra thick] (0:-1.25) arc (0:-45:-1.25);
    \node[fill, red, circle,inner sep=2pt,label={}] at (0,-1.25){};
    \node[fill, red, circle,inner sep=2pt,label={}] at (0,1.25){};

    \end{axis}       
\end{tikzpicture}
\end{center}
\vskip -5mm
\caption{Sketch of all possible initial conditions with the same initial energy $E_0$ in the $(\omega_0x_0,v_0)$ coordinate system. 
Square of the coordinates corresponds to initial potential energy $E_{0P}=\omega_0^2x_0^2$ and initial kinetic energy $E_{0K}=v_0^2$ respectively. 
This representation gives us a useful visualization, e.g.: all initial conditions with $E_{0P}>E_{0K}$ are represented by two arcs, i.e. points with $\theta\in(-\pi/4,\pi/4)\cup(3\pi/4,5\pi/4)$ (blue dotted arcs); initial conditions with $E_{0K}=E_0$ and $E_{0P}=0$ are represented by two points on a circle with $\theta=\lbrace \pi/2,3\pi/2\rbrace$ (two red filled circles); etc.  }
\label{fig:skica}
\end{figure}
Using \eqref{polar} in \eqref{Et1} and \eqref{Etc}, we obtain the energy of the under-damped and over-damped regime  
\begin{equation}
E(t, \theta)=E_0e^{-2\gamma t}\left( \cos^2(\omega t)+\gamma\cos2\theta\frac{\sin(2\omega t)}{\omega}+\left(\omega_0^2+\gamma^2+2\omega_0\gamma\sin2\theta\right)\frac{\sin^2(\omega t)}{\omega^2}\right)\,,
\label{Et2}
\end{equation}
and the energy of the critically damped regime 
\begin{equation}
E_c(t, \theta)=E_0e^{-2\omega_0 t}\left( 1+2\omega_0(\cos2\theta)t+2\omega_0^2\left(1+\sin2\theta\right)t^2\right)\,,
\label{Et2c}
\end{equation}
as functions of $\theta$, instead of $x_0$ and $v_0$. Now we integrate energy \eqref{Et2} over all time, i.e.
\begin{equation}
I(\gamma,\theta)=\int_0^{\infty}E(t)dt\,,
\label{Int1}
\end{equation}
%
and obtain
\begin{equation}
I(\gamma,\theta)=\frac{E_0}{2\omega_0}\left(\frac{\omega_0^2+\gamma^2}{\gamma\omega_0}+\frac{\gamma}{\omega_0}\cos2\theta+\sin2\theta\right)\,.
\label{Int12}
\end{equation}
Integral \eqref{Int12} is valid for all three regimes, i.e. for any $\gamma>0$. 

We note here that the energy (see \eqref{Et1} and \eqref{Etc}) is invariant to a simultaneous change of the signs of the initial conditions, i.e. to the change $(x_0, v_0)\rightarrow(-x_0, -v_0)$ (but not to $x_0\rightarrow -x_0$ or $v_0\rightarrow -v_0$ separately). This change of signs corresponds to the change in angle $\theta\rightarrow\theta+\pi$, therefore, functions $\eqref{Et2}$, \eqref{Et2c} and \eqref{Int12} are all periodic in $\theta$ with period $\pi$.

\begin{figure}
\begin{center}
\begin{tikzpicture}
        \begin{axis}[
        width=0.485\textwidth,
        height=0.3\textwidth,
        xmin = 0,
        xmax = 3,
        ymin = 0,
        ymax = 5.5,
        xtick={0.1,0.71,1,2,3},
        ytick={0,1,...,5},
        every tick label/.append style={font=\small},
        ylabel near ticks,
        xlabel near ticks,
        xlabel = {\small $\gamma/\omega_0$},
        ylabel = {\small $I\left(\gamma,\theta\right)\omega_0/E_0$},
        legend entries = {{\footnotesize $\theta=0$},{\footnotesize $\theta=\pi/4$},{\footnotesize $\theta=\pi/2$}}
        ]
        \addplot [smooth,domain=0.09:3,samples=100, very thick, red](\x,{0.5*(1+2*\x^2)/\x});
        \addplot [dotted,domain=0.09:3,samples=100, very thick, black](\x,{0.5*(1+\x^2+\x)/\x});
        \addplot [dashed,domain=0.09:3,samples=100, very thick, blue](\x,{0.5*(1)/\x});

        
        \draw[->,>=stealth] (axis cs: 0.707,0) -- (axis cs: 0.707,1.4);
        \draw[->,>=stealth] (axis cs: 1,0) -- (axis cs: 1,1.5);
                
        \end{axis}  
    \end{tikzpicture}
\end{center}
\caption{Integral \eqref{Int12} for three initial conditions $\theta=\lbrace 0, \pi/4, \pi/2\rbrace$. 
}
\label{fig:integral1}
\end{figure}
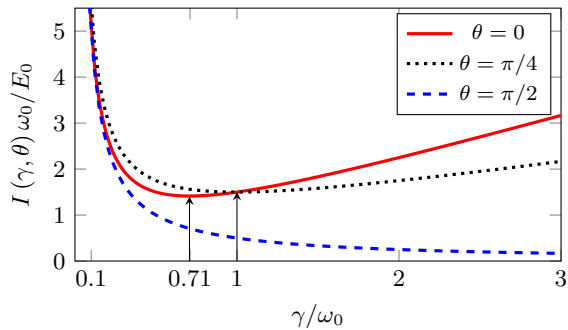

In Fig.\ \ref{fig:integral1} we show the integral \eqref{Int12} for $\gamma\in\left[0.1\omega_0,3\omega_0\right]$ for three different initial conditions, i.e. for $\theta=\lbrace{0, \pi/4, \pi/2\rbrace}$. We can see that $I(\gamma, \theta=0)$ (red solid curve), with purely potential initial energy and zero initial kinetic energy, attains minimum for $\gamma=0.707\omega_0$ (rounded to three decimal places), i.e. well in the under-damped regime. For the initial condition with equal potential and kinetic energy, $I(\gamma, \theta=\pi/4)$ (black dotted curve) attains minimum for $\gamma=\omega_0$, i.e. at the critical damping condition. Interestingly, for the initial condition with purely kinetic energy and zero potential energy, $I(\gamma, \theta=\pi/2)$ (blue dashed curve) has no minimum in the displayed range of damping coefficients, therefore here we explicitly show this function 
\begin{equation}
I(\gamma,\theta=\pi/2)=\frac{E_0}{2\gamma}\,,
\label{Int123}
\end{equation}
and it is clear that \eqref{Int123} 
has no minimum. This is easy to understand from a physical point of view, i.e. if all the initial energy is kinetic, the higher the damping coefficient, the faster the energy dissipation will be. 

If we consider the optimal damping as the one for which the integral \eqref{Int12} is minimal, we can easily determine the optimal damping coefficient $\gamma_{\opt}(\theta)$ from the condition 
\begin{equation}
\frac{\partial I(\gamma, \theta)}{\partial\gamma}\bigg|_{\gamma_{\opt}}=0\,,
\label{1min}
\end{equation}
and we obtain
\begin{equation}
\gamma_{\opt}(\theta)=\sqrt{\frac{1}{2\cos^2\theta}}\omega_0\,.
\label{gopt}
\end{equation}

In Fig.\ \ref{fig:opt1} we show the optimal damping coefficient \eqref{gopt} for $\theta\in\left[0,2\pi\right]$ (function \eqref{gopt} has a period $\pi$, but here we choose this interval for completeness), and here we comment on the shown results with respect to the relationship between initial potential energy ($E_{0P}=\omega_0^2x_0^2$) and initial kinetic energy ($E_{0K}=v_0^2$) for any given initial condition, i.e. for any $\theta$:
\begin{itemize}
\item Initial conditions with $E_{0P}>E_{0K}$ correspond to the set $\theta\in\left(-\pi/4,\pi/4\right)\cup\left(3\pi/4,5\pi/4\right)$. For these initial conditions, optimal damping coefficients \eqref{gopt} are in the under-damped regime, i.e. $\gamma_{\opt}\in\left[\sqrt{2}\omega_0/2,\omega_0\right)$, with the minimum value $\gamma_{\opt}=\sqrt{2}\omega_0/2=0.707\omega_0$ (rounded to three decimal places) obtained for $\theta=\lbrace0,\pi\rbrace$, i.e. for two initial conditions with $E_0=E_{0P}$ and $E_{0K}=0$.
\item Initial conditions with $E_{0P}=E_{0K}$ correspond to four points $\theta=\lbrace\pi/4,3\pi/4, 5\pi/4, 7\pi/4\rbrace$ with optimal damping coefficient \eqref{gopt} equal to critical damping, i.e. $\gamma_{\opt}=\omega_0$. 
\item Initial conditions with $E_{0K}>E_{0P}$ correspond to the set $\theta\in\left(\pi/4,3\pi/4\right)\cup\left(5\pi/4,7\pi/4\right)$. For these initial conditions, optimal damping coefficients \eqref{gopt} are in the over-damped regime, i.e. $\gamma_{\opt}\in\left(\omega_0,\infty\right)$, where $\gamma_{\opt}$ diverges for $\theta=\lbrace\pi/2,3\pi/2\rbrace$, i.e. for two initial conditions with $E_{0K}=E_0$ and $E_{0P}=0$.
\end{itemize}
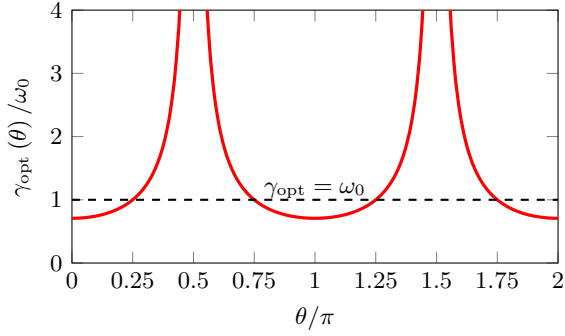
\begin{figure}
\begin{center}
\begin{tikzpicture}
        \begin{axis}[
        width=0.485\textwidth,
        height=0.3\textwidth,
        xmin = 0,
        xmax = 2,
        ymin = 0,
        ymax = 4,
        xtick={0,0.25,0.5,0.75,1,...,2},
        ytick={0,1,...,4},
        every tick label/.append style={font=\small},
        ylabel near ticks,
        xlabel near ticks,
        xlabel = {\small $\theta/\pi$},
        ylabel = {\small $\gamma_{\opt}\left(\theta\right)/\omega_0$},
        ]
        \addplot [smooth,domain=0:2,samples=100, very thick, red](\x,{(1+cos(2*\x*3.1415 r))^(-0.5)});
        \draw[-,>=stealth, thick, dashed] (axis cs: 0,1) -- (axis cs: 2,1);
        \node[] at (axis cs: 1,1.15) {\small $\gamma_{\opt}=\omega_0$};

        \end{axis}  
    \end{tikzpicture}
\end{center}
\caption{Optimal damping coefficient \eqref{gopt} (solid red curve) as a function of all possible initial conditions, i.e. for $\theta\in\left[0,2\pi\right]$. Below the dashed horizontal line, optimal damping coefficients are in the under-damped regime, above the line in the over-damped regime, and in the critically damped regime at the crossing points of the line and the solid red curve.  }
\label{fig:opt1}
\end{figure}

Before closing this subsection, we would like to point out two more ways in which we can write relation \eqref{gopt} that will prove useful when dealing with MDOF systems. The ratio of the initial potential energy to the initial total energy is
\begin{equation}
\beta=\frac{E_{0P}}{E_0}=\cos^2\theta\,,
\label{ratio1D}
\end{equation}
where we used first of the relations \eqref{polar} and $E_{0P}=\omega_0^2x_0^2$. Using \eqref{ratio1D}, optimal damping coefficient \eqref{gopt} can be written as a function of the fraction of potential energy in the initial total energy, i.e.
\begin{equation}
\gamma_{\opt}(\beta)=\sqrt{\frac{1}{2\beta}}\omega_0\,.
\label{gopt_pot}
\end{equation}
Thus, from \eqref{gopt_pot} one can simply see that $\gamma_{\opt}$ is in the under-damped regime for $\beta\in(1/2,1]$, in the critically damped regime for $\beta=1/2$ and in the over-damped regime for $\beta\in[0,1/2)$. Using $\beta=\omega_0^2x_0^2/E_0$ in \eqref{gopt_pot} we can express the optimal damping coefficient in yet another way, as a function of the initial displacement $x_0$, i.e. 
\begin{equation}
\gamma_{\opt}(x_0)=\sqrt{\frac{E_0}{2x_0^2}}=\sqrt{\frac{v_0^2+\omega_0^2x_0^2}{2x_0^2}}\,,
\label{gopt_x0}
\end{equation}
where $x_0\in[-\sqrt{E_0}/\omega_0,\sqrt{E_0}/\omega_0]$ and for $v_0$ the condition $v_0^2=E_0-\omega_0^2x_0^2$ holds. One of the benefits of relation \eqref{gopt_x0} is that it can be seen most directly that the optimal damping coefficient does not distinguish initial conditions $(\pm x_0,\pm v_0)$ and $(\pm x_0,\mp v_0)$, which is a shortcoming of this optimization criterion, because the energy as a function of time is not the same for those two types of initial conditions (see \eqref{Et1} and \eqref{Etc}) and the energy decay may differ significantly depending on which of those initial conditions is in question. We will deal with these and other issues of energy integral minimization as an optimal damping criterion in the subsection \ref{issues}.

\subsection{Minimization of the energy integral averaged over a set of initial conditions and optimal damping in dependence of the chosen set} 
\label{1Daverage}

\begin{figure}
\begin{center}
\begin{tikzpicture}
        \begin{axis}[
        width=0.485\textwidth,
        height=0.3\textwidth,
        xmin = 0,
        xmax = 3,
        ymin = 0,
        ymax = 5,
        xtick={0.1,0.78,1,1.66,2,3,...,5},
        ytick={0,1,...,5},
        every tick label/.append style={font=\small},
        ylabel near ticks,
        xlabel near ticks,
        xlabel = {\small $\gamma/\omega_0$},
        ylabel = {\small $\overline{I}\left(\gamma,\phi_1,\phi_2\right)\omega_0/E_0$},
        legend entries = {{\footnotesize $\phi_1=-\pi/4$\,, $\phi_2=\pi/4$},{\footnotesize $\phi_1=\pi/4$\,, $\phi_2=3\pi/4$},{\footnotesize $\phi_1=-\pi/4$\,, $\phi_2=3\pi/4$}}
        ]
        \addplot [smooth,domain=0.1:3,samples=100, very thick, red](\x,{0.5*((1+\x^2)/\x+\x*(sin(2*3.1415/4 r)+sin(2*3.1415/4 r))/3.1415)});
        \addplot [dashed,domain=0.1:3,samples=100, very thick, blue](\x,{0.5*((1+\x^2)/\x+\x*(-2)/3.1415)});
        \addplot [dotted,domain=0.1:3,samples=100, very thick, black](\x,{0.5*((1+\x^2)/\x)});
         
        \draw[->,>=stealth] (axis cs: 0.78,0) -- (axis cs: 0.78,1.3);
        \draw[->,>=stealth] (axis cs: 1,0) -- (axis cs: 1,1);
        \draw[->,>=stealth] (axis cs: 1.66,0) -- (axis cs: 1.66,0.625);
                
        \end{axis}  
    \end{tikzpicture}
\end{center}
\caption{Averaged integral \eqref{Int22} for three sets of initial conditions. 
}
\label{fig:int2}
\end{figure}
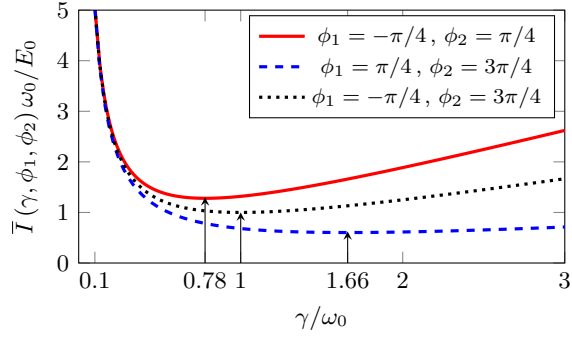

Now we calculate the average of the integral \eqref{Int1} over a set of initial conditions with $\theta\in\left[\phi_1,\phi_2\right]$, i.e.
\begin{equation}
\overline{I}(\gamma,\phi_1, \phi_2)=\frac{1}{\phi_2-\phi_1}\int_{\phi_1}^{\phi_2}I(\gamma,\theta)d\theta\,,
\label{Int2}
\end{equation}
and we obtain
\begin{equation}
\overline{I}(\gamma,\phi_1, \phi_2)=\frac{E_0}{2\omega_0}\left(\frac{\omega_0^2+\gamma^2}{\gamma\omega_0}+\frac{\gamma}{2\omega_0(\phi_2-\phi_1)}(\sin2\phi_2-\sin2\phi_1)+\frac{1}{2(\phi_2-\phi_1)}(\cos2\phi_1-\cos2\phi_2)\right)\,.
\label{Int22}
\end{equation}
In Fig.\ \ref{fig:int2} we show averaged integral \eqref{Int22} for three different sets of initial conditions. For the set of initial conditions with $\phi_1=-\pi/4$ and $\phi_2=\pi/4$, i.e. with $E_{0P}\geq E_{0K}$ (where the equality holds only at the end points of the set), minimum of the averaged integral (solid red curve) is at $\gamma=0.781\omega_0$ (rounded to three decimal places). For the set of initial conditions with $\phi_1=\pi/4$ and $\phi_2=3\pi/4$, i.e. with $E_{0K}\geq E_{0P}$ (where the equality holds only at the end points of the set), minimum of the averaged integral (dashed blue curve) is at $\gamma=1.658\omega_0$ (rounded to three decimal places). For the set of mixed initial conditions with $\phi_1=-\pi/4$ and $\phi_2=3\pi/4$, i.e. with $E_{0P}>E_{0K}$ and $E_{0K}>E_{0P}$ points equally present in the set, minimum of the averaged integral (dotted black curve) is at the critical damping condition $\gamma=\omega_0$.

If we consider the optimal damping as the one for which the averaged integral \eqref{Int22} is minimal, we can easily determine the optimal damping coefficient $\overline{\gamma}_{\opt}(\phi_1,\phi_2)$ form the condition 
\begin{equation}
\frac{\partial \overline{I}(\gamma, \phi_1, \phi_2)}{\partial\gamma}\bigg|_{\overline{\gamma}_{\opt}}=0\,,
\label{2min}
\end{equation}
and we obtain
\begin{equation}
\overline{\gamma}_{\opt}(\phi_1,\phi_2)=\sqrt{\frac{2(\phi_2-\phi_1)}{2(\phi_2-\phi_1)+\sin2\phi_2-\sin2\phi_1}}\omega_0\,.
\label{gopt2}
\end{equation}
We note here that averaged integral \eqref{Int22} and optimal damping coefficient \eqref{gopt2} are not periodic functions in variables $\phi_1$ and $\phi_2$, if we keep one variable fixed and change the other. But they are periodic, with period $\pi$, if we change both variables simultaneously.   

In Fig.\ \ref{fig:opt2} we show the optimal damping coefficient \eqref{gopt2} as a function of $\phi_2$ with fixed $\phi_1=0$, and the results shown can be summarized as follows:
\begin{itemize}
\item For $\phi_1=0$ and $\phi_2\in\left[0,\pi/2\right)\cup\left(\pi,3\pi/2\right)$, the optimal damping coefficient \eqref{gopt2} is in the under-damped regime. In this case, integral \eqref{Int2} is averaged over sets that have more points corresponding to initial conditions with $E_{0P}>E_{0K}$, in comparison to the points corresponding to initial conditions with $E_{0K}>E_{0P}$.
\item For $\phi_1=0$ and $\phi_2=\lbrace \pi/2, \pi, 3\pi/2, 2\pi \rbrace$, the optimal damping coefficient \eqref{gopt2} is equal to critical damping. In this case, integral \eqref{Int2} is averaged over sets that have equal amount of points corresponding to initial conditions with $E_{0P}>E_{0K}$ and initial conditions with $E_{0K}>E_{0P}$.
\item For $\phi_1=0$ and $\phi_2\in\left(\pi/2,\pi\right)\cup\left(3\pi/2,2\pi\right)$, the optimal damping coefficient \eqref{gopt2} is in the over-damped regime. In this case, integral \eqref{Int2} is averaged over sets that have more points corresponding to initial conditions with $E_{0K}>E_{0P}$, in comparison to the points corresponding to initial conditions with $E_{0P}>E_{0K}$.
\end{itemize}
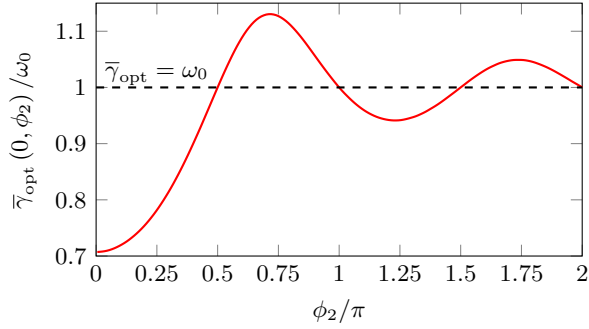
\begin{figure}
\begin{center}
\begin{tikzpicture}
        \begin{axis}[
        width=0.485\textwidth,
        height=0.3\textwidth,
        xmin = 0,
        xmax = 2,
        ymin = 0.7,
        ymax = 1.15,
        xtick={0,0.25,0.5,0.75,1,...,2},
        ytick={0.7,0.8,...,1.1},
        every tick label/.append style={font=\small},
        ylabel near ticks,
        xlabel near ticks,
        xlabel = {\small $\phi_2/\pi$},
        ylabel = {\small $\overline{\gamma}_{\opt}\left(0,\phi_2\right)/\omega_0$},
        ]
        \addplot [smooth,domain=0.005:2,samples=100, thick, red](\x,{((2*\x*3.1415)^(0.5))*(2*\x*3.1415+sin(2*\x*3.1415 r))^(-0.5)});
        \draw[-,>=stealth, thick, dashed] (axis cs: 0,1) -- (axis cs: 2,1);
            
        \node[] at (axis cs: 0.25,1.025) {\small $\overline{\gamma}_{\opt}=\omega_0$};

        \end{axis}  
    \end{tikzpicture}
\end{center}
\caption{Optimal damping coefficient \eqref{gopt2} (solid red curve) as a function of $\phi_2\in\left[0,2\pi\right]$ for fixed $\phi_1=0$. Below the dashed horizontal line, optimal damping coefficients are in the under-damped regime, above the line in the over-damped regime, and in the critically damped regime at the crossing points of the line and the solid red curve.  }
\label{fig:opt2}
\end{figure}

\section{2-DOF systems with MPD}
\label{2Dsection}


\begin{figure}[h!t!]
\begin{center}
\includegraphics[width=0.55\textwidth]{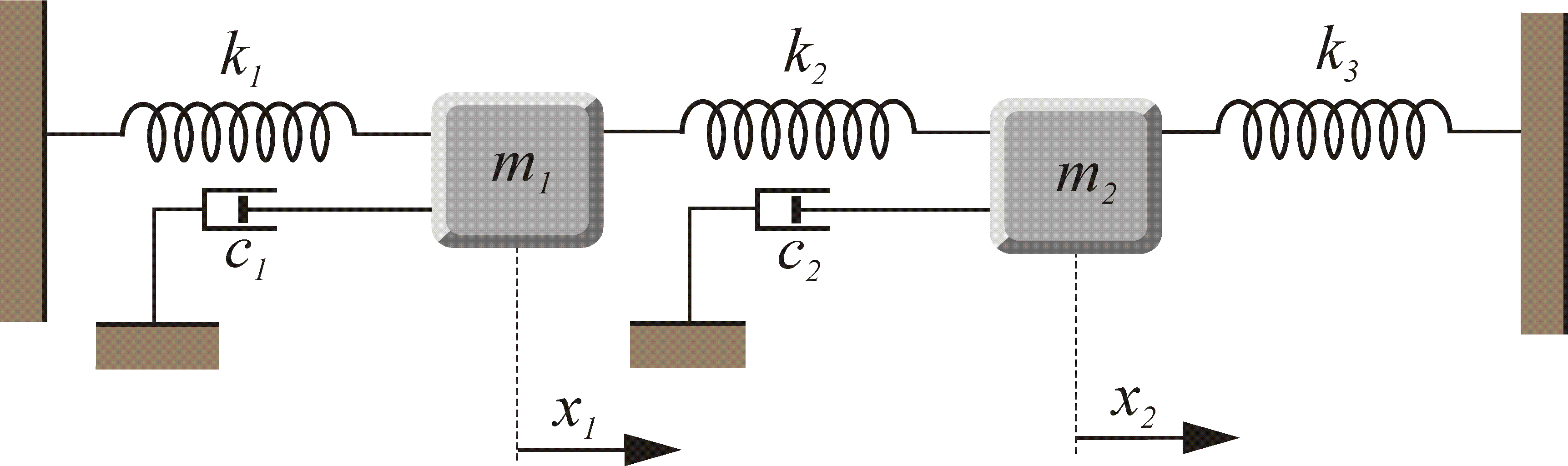}
\end{center}
\caption{Schematic figure of a 2-DOF system.  }
\label{fig:skica2D}
\end{figure}

Here we consider 2-DOF system shown schematically in Fig.\ \ref{fig:skica2D}. The corresponding equations of motion are
\begin{equation}
\begin{split}
m_1\ddot x_1(t)=-c_1\dot x_1(t)-k_1x_1(t)-k_2\left(x_1(t)-x_2(t)\right)\,,
\\m_2\ddot x_2(t)=-c_2\dot x_2(t)-k_3x_2(t)+k_2\left(x_1(t)-x_2(t)\right)\,.
\label{eq2D}
\end{split}
\end{equation}
%
We will consider MPD \cite{muravyov1998geometrical}, i.e. masses $\lbrace m_1,m_2\rbrace$, spring constants $\lbrace k_1,k_2,k_3\rbrace$, and dampers $\lbrace c_1,c_2\rbrace$ can in general be mutually different but the condition $c_1/m_1=c_2/m_2$ holds. In this case we can use modal analysis \cite{fu2001modal, grigoriu2021linear} and the system of equations \eqref{eq2D} can be written via modal coordinates \cite{grigoriu2021linear} as
\begin{equation}
\begin{split}
\ddot q_1(t)+2\gamma\dot q_1(t)+\omega_{01}^2q_1(t)=0\,
\\\ddot q_2(t)+2\gamma\dot q_2(t)+\omega_{02}^2q_2(t)=0\,,
\label{eq2Dnorm}
\end{split}
\end{equation}
where $q_i(t)$ and $\omega_{0i}$, with $i=\lbrace 1,2\rbrace$, denote the modal coordinates and undamped modal frequencies of the two modes, while $\gamma=c_i/2m_i$ is the damping coefficient. In the analysis that we will carry out in this subsection, we will not need the explicit connection of modal coordinates $q_i(t)$ and mass coordinates, i.e. displacements $x_i(t)$, and we will deal with this in the next subsection when considering a specific example with given masses, springs and dampers. Similarly as in Section \ref{1Dsection} (see \eqref{xud}), the general solution for the $i$-th mode can be written as
\begin{equation}
q_i(t)=e^{-\gamma t}\left(q_{0i}\cos(\omega_i t)+\frac{\dot q_{0i}+\gamma q_{0i}}{\omega_i}\sin(\omega_i t)\right)\,,
\label{mod2D}
\end{equation}
where $\omega_i=\sqrt{\omega_{0i}^2-\gamma^2}$ is the damped modal frequency, and $q_i(0)\equiv q_{0i}$ and $\dot q_i(0)\equiv \dot q_{0i}$ are the initial conditions of the $i$-th mode. Thus, the reasoning and the results presented in Section \ref{1Dsection}, with some adjustments, can by applied for the analysis of the 2-DOF system we are considering here.

The energy of the system is
\begin{equation}
E(t)=\sum_{i=1}^2\frac{m_i\dot{x}_i(t)^2}{2}+\frac{k_1x_1(t)^2}{2}+\frac{k_3x_2(t)^2}{2}+\frac{k_2(x_1(t)-x_2(t))^2}{2}\,,
\label{Emass2D}
\end{equation}
and we take that the modal coordinates are normalised so that \eqref{Emass2D} can be written as  
\begin{equation}
E(t)=\sum_{i=1}^2E_i(t)=\sum_{i=1}^2\left(\dot q_i(t)^2+\omega_{0i}^2q_i(t)^2\right)\,
\label{energy2D}
\end{equation}
where $E_i(t)$ in \eqref{energy2D} denotes the energy of the $i$-th mode. Total energy at $t=0$, i.e. the initial energy, is given by 
\begin{equation}
E_0=\sum_{i=1}^2E_{0i}=\sum_{i=1}^2\left(E_{0Ki}+E_{0Pi}\right)=\sum_{i=1}^2\left(\dot q_{0i}^2+\omega_{0i}^2q_{0i}^2\right)\,,
\label{energy02D}
\end{equation}
where $E_{0i}$ denotes the initial energy of the $i$-th mode, $E_{0Ki}=\dot q_{0i}^2$ and $E_{0Pi}=\omega_{0i}^2q_{0i}^2$ denote initial kinetic and initial potential energy of the $i$-th mode.

All possible initial conditions with the same initial energy \eqref{energy02D} can be expressed similarly as in the SDOF case (see \eqref{polar} and Fig.\ \ref{fig:skica}) but with two pairs of polar coordinates, one pair for each mode. For the $i$-th mode we have radius $r_i=\sqrt{E_{0i}}$ and angle $\theta_i=\arctan\left(\frac{\dot q_{0i}}{\omega_{0i}q_{0i}}\right)$, i.e. we can write   
\begin{equation}
\begin{split}
\omega_{0i}q_{0i}=r_i\cos\theta_i\\
\dot q_{0i}=r_i\sin\theta_i\,.
\label{polar2D}
\end{split}
\end{equation}
Thus, each initial condition with energy $E_0=E_{01}+E_{02}$ can be represented by points on two circles with radii $r_1=\sqrt{E_{01}}$ and $r_2=\sqrt{E_{02}}$, for which condition $r_1^2+r_2^2=E_0$ holds, and with angles $\theta_1$ and $\theta_2$ that tell us how initial potential and initial kinetic energy are distributed within the modes. Using relation \eqref{Et2} for SDOF systems, we can write the energy of the $i$-th mode in polar coordinates \eqref{polar2D} as
\begin{equation}
E_i(t)=E_{0i}e^{-2\gamma t}\left( \cos^2(\omega_i t)+\gamma\cos2\theta_i\frac{\sin(2\omega_i t)}{\omega_i}+\left(\omega_{0i}^2+\gamma^2+2\omega_{0i}\gamma\sin2\theta_i\right)\frac{\sin^2(\omega_i t)}{\omega_i^2}\right)\,
\label{Eipolar}
\end{equation}
for the under-damped ($\gamma<\omega_{0i}$) and over-damped ($\gamma>\omega_{0i}$) regime, and the energy of the $i$-th mode in the critically damped regime is obtained analogously using the relation \eqref{Et2c}.

Consequently, the integral of the energy \eqref{energy2D} over the entire time, for some arbitrary initial condition, is simply calculated using relation \eqref{Int12} for each individual mode, we obtain
\begin{equation}
I(\gamma, \lbrace E_{0i}\rbrace, \lbrace \theta_i\rbrace)=\sum_{i=1}^2\int_0^{\infty}E_i(t)dt=\sum_{i=1}^2\frac{E_{0i}}{2\omega_{0i}}\left(\frac{\omega_{0i}^2+\gamma^2}{\gamma\omega_{0i}}+\frac{\gamma}{\omega_{0i}}\cos2\theta_i+\sin2\theta_i\right)\,.
\label{int2D1}
\end{equation}
Furthermore, initial energy of the $i$-th mode can be written as $E_{0i}=a_i^2E_0$, where coefficient $a_i^2\in[0,1]$ denotes the fraction of the initial energy of the $i$-th mode in the total initial energy. Coefficients of the two modes satisfy $a_1^2+a_2^2=1$ and therefore can be parameterized as
\begin{equation}
\begin{split}
a_1=\cos\psi\\
a_2=\sin\psi,
\label{ai2D}
\end{split}
\end{equation}
where $\psi\in[0,\pi/2]$. Taking \eqref{ai2D} into account, we can write \eqref{int2D1} as  
\begin{equation}
I(\gamma, \psi, \theta_1, \theta_2)=E_0\sum_{i=1}^2\frac{a_i^2}{2\omega_{0i}}\left(\frac{\omega_{0i}^2+\gamma^2}{\gamma\omega_{0i}}+\frac{\gamma}{\omega_{0i}}\cos2\theta_i+\sin2\theta_i\right)\,.
\label{int2D12}
\end{equation}
If we consider the optimal damping coefficient as the one for which the integral \eqref{int2D12} is minimal, we can easily determine the optimal damping coefficient form the condition 
\begin{equation}
\frac{\partial I(\gamma, \psi, \theta_1, \theta_2)}{\partial\gamma}\bigg|_{\gamma_{\opt}}=0\,,
\label{1min2D}
\end{equation}
and we obtain
\begin{equation}
\gamma_{\opt}(\psi,\theta_1,\theta_2)=\sqrt{\frac{\omega_{01}^2\omega_{02}^2}{2\omega_{02}^2\cos^2\psi\cos^2\theta_1+2\omega_{01}^2\sin^2\psi\cos^2\theta_2}}\,.
\label{gammaopt}
\end{equation}
It is easy to see that, for any fixed $\psi$, the function \eqref{gammaopt} has smallest magnitude for $\cos^2\theta_1=\cos^2\theta_2=1$, which corresponds to the initial conditions with initial energy comprised only of potential energy distributed within the two modes, i.e $E_0=E_{0P1}+E_{0P2}$. In that case we can write the denominator of \eqref{gammaopt} as
\begin{equation}
f(\psi)=\sqrt{2\omega_{02}^2\cos^2\psi+2\omega_{01}^2\sin^2\psi}=\sqrt{2(\omega_{02}^2-\omega_{01}^2)\cos^2\psi+2\omega_{01}^2}\,,
\label{denominator}
\end{equation}
where we used $\sin^2\psi=1-\cos^2\psi$. Since $\omega_{01}<\omega_{02}$, the function \eqref{denominator} has maximum for $\psi=0$. Thus, the minimum value of the optimal damping coefficient \eqref{gammaopt} is $\sqrt{2}\omega_{01}/2$, and it is obtained for $\psi=0$ and $\theta_1=\lbrace0,\pi\rbrace$, which corresponds to the initial conditions with initial energy comprised only of potential energy in the first mode, i.e. $E_0=E_{0P1}$. On the other hand, for any fixed $\psi$, the function \eqref{gammaopt} has singularities for $\cos^2\theta_1=\cos^2\theta_2=0$, which corresponds to the initial conditions with initial energy comprised only of kinetic energy. Thus, the range of the optimal damping coefficient \eqref{gammaopt} is
\begin{equation}
    \gamma_{\opt}\in\left[\sqrt{2}\omega_{01}/2,+\infty\right)\,.
    \label{rangegamma}
\end{equation}
Now we calculate the average of the integral \eqref{int2D12} over a set of all initial conditions, we obtain
\begin{equation}
\overline{I}(\gamma)=\frac{1}{2\pi^3}\int_0^{\pi/2}d\psi\int_0^{2\pi}d\theta_1\int_0^{2\pi}d\theta_2\,I(\gamma, \psi, \theta_1, \theta_2)=\frac{E_0}{4}\sum_{i=1}^2\left(\frac{\omega_{0i}^2+\gamma^2}{\gamma\omega_{0i}^2}\right)\,,
\label{int2D2}
\end{equation}
and from the condition
\begin{equation}
\frac{\partial \overline{I}(\gamma)}{\partial\gamma}\bigg|_{\overline{\gamma}_{\opt}}=0\,,
\label{2min2D}
\end{equation}
we find  that the optimal damping coefficient with respect to the averaged integral \eqref{int2D2} is given by
\begin{equation}
\overline{\gamma}_{\opt}=\sqrt{\frac{2\omega_{01}^2\omega_{02}^2}{\omega_{01}^2+\omega_{02}^2}}\,.
\label{avgammaopt}
\end{equation}

In order to more easily analyze the behavior of the damping coefficient \eqref{gammaopt} with regard to the distribution of the initial potential energy within the modes and its relationship with the damping coefficient \eqref{avgammaopt}, similarly as in subsection \ref{1Dinitialcond} (see \eqref{ratio1D} and \eqref{gopt_pot}), we define the ratio of the initial potential energy of the $i$-th mode and the total initial energy, i.e.
\begin{equation}
\beta_i=\frac{E_{0Pi}}{E_0}\,.
\label{beta1}
\end{equation}
Since the initial potential energy satisfies $E_{0P}=E_{0P1}+E_{0P2}\leq E_0$, we have $\beta_i\in[0,1]$ and the condition $0\leq(\beta_1+\beta_2)\leq 1$ holds. Taking $E_{0Pi}=E_{0i}\cos^2\theta_i$ (see \eqref{polar2D}) and $E_{0i}=a_i^2E_0$ with \eqref{ai2D} into account, we have 
\begin{equation}
\begin{split}
\beta_1=\cos^2\psi\cos^2\theta_1\\
\beta_2=\sin^2\psi\cos^2\theta_2.
\label{beta2}
\end{split}
\end{equation}
Using \eqref{beta2}, relation \eqref{gammaopt} can be written as
\begin{equation}
\gamma_{\opt}(\beta_1,\beta_2)=\sqrt{\frac{\omega_{01}^2\omega_{02}^2}{2\omega_{02}^2\beta_1+2\omega_{01}^2\beta_2}}\,.
\label{gammaopt2}
\end{equation}
For clarity, we will repeat briefly, the minimum value of \eqref{gammaopt2} is $\sqrt{2}\omega_{01}/2$, obtained for $\beta_1=1$ and $\beta_2=0$ (or in terms of the angles in \eqref{gammaopt}, for $\psi=0$ and $\theta_1=\lbrace 0,\pi\rbrace$), while $\gamma_{\opt}\rightarrow+\infty$ for $\beta_1=\beta_2=0$ (or in terms of the angles in \eqref{gammaopt}, for any $\psi$ with $\theta_1=\lbrace \pi/2,3\pi/2\rbrace$ and $\theta_2=\lbrace \pi/2,3\pi/2\rbrace$). The benefit of relation \eqref{gammaopt2} is that we expressed \eqref{gammaopt} through two variables instead of three, i.e.\ this way we lost information about the signs of the initial conditions and about distribution of initial kinetic energy within the modes, but the optimal damping coefficient \eqref{gammaopt} does not depend on those signs anyway, due to the squares of trigonometric functions in variables $\theta_1$ and $\theta_2$, and, for a fixed distribution of initial potential energy within the modes, the optimal damping coefficient \eqref{gammaopt} is constant for different distributions of initial kinetic energy within the modes. By looking at relations \eqref{gammaopt2} and \eqref{avgammaopt}, it is immediately clear that $\gamma_{\opt}(\beta_1,\beta_2)=\overline{\gamma}_{\opt}$ for
\begin{equation}
    \omega_{02}^2\beta_1+\omega_{01}^2\beta_2=\frac{\omega_{01}^2+\omega_{02}^2}{4},
    \label{conditonBeta}
\end{equation}
while $\gamma_{\opt}(\beta_1,\beta_2)<\overline{\gamma}_{\opt}$ if the left hand side of relation \eqref{conditonBeta} is greater than the right hand side, and $\gamma_{\opt}(\beta_1,\beta_2)>\overline{\gamma}_{\opt}$ if the left hand side of relation \eqref{conditonBeta} is smaller than the right hand side.

Again, similarly as in subsection \ref{1Dinitialcond} (see \eqref{gopt_x0}), using $\beta_i=\omega_{0i}^2q_{0i}^2/E_0$ we can express the optimal damping coefficient \eqref{gammaopt2} as a function of the initial modal coordinates as well, i.e.
\begin{equation}
\gamma_{\opt}(q_{01},q_{02})=\sqrt{\frac{E_0}{2q_{01}^2+2q_{02}^2}}\,,
\label{gammaopt_q0}
\end{equation}
where $q_{0i}\in[-\sqrt{E_0}/\omega_{0i},\sqrt{E_0}/\omega_{0i}]$ and the condition $0\leq (\omega_{01}^2q_{01}^2+\omega_{02}^2q_{02}^2)\leq E_0$ holds. We can express condition \eqref{conditonBeta} in terms of initial modal coordinates, i.e. $\gamma_{\opt}(\lbrace q_{0i}\rbrace)=\overline{\gamma}_{\opt}$ for
\begin{equation}
    \frac{q_{01}^2+q_{02}^2}{E_0}=\frac{\omega_{01}^2+\omega_{02}^2}{4\omega_{01}^2\omega_{02}^2},
    \label{conditionQ}
\end{equation}
while $\gamma_{\opt}(\lbrace q_{0i}\rbrace)<\overline{\gamma}_{\opt}$ if the left hand side of relation \eqref{conditionQ} is greater than the right hand side, and $\gamma_{\opt}(\lbrace q_{0i}\rbrace)>\overline{\gamma}_{\opt}$ if the left hand side of relation \eqref{conditionQ} is smaller than the right hand side.   

We note here that we did not use explicit values of the undamped modal frequencies $\omega_{01}$ and $\omega_{02}$ in the analysis so far, and relations presented so far are valid for any 2-DOF system with MPD. In the next subsection, we provide a more detailed quantitative analysis using an example with specific values of modal frequencies.    

\subsection{Quantitative example}
\label{example2D}

Here we consider the 2-DOF system as the one shown schematically in Fig.\ \ref{fig:skica2D}, but with $m_1=m_2=m$, $k_1=k_2=k_3=k$ and $c_1=c_2=c$. The corresponding equations of motion are
\begin{equation}
\begin{split}
m\ddot x_1(t)=-c\dot x_1(t)-kx_1(t)-k\left(x_1(t)-x_2(t)\right)\,,
\\m\ddot x_2(t)=-c\dot x_2(t)-kx_2(t)+k\left(x_1(t)-x_2(t)\right)\,.
\label{eq2D1}
\end{split}
\end{equation}
For completeness, we will investigate here the behavior of the optimal damping coefficient given by the minimization of the energy integral for different initial conditions, and its relationship with the optimal damping coefficient given by the minimization of the averaged energy integral, in all three coordinate systems that we introduced in the previous subsection and additionally in the coordinate system defined by the initial displacements of the masses. System of equations \eqref{eq2D1} can be easily recast to the form \eqref{eq2Dnorm} with the modal coordinates
\begin{equation}
\begin{split}
    q_1(t)=\sqrt{\frac{m}{4}}\left(x_1(t)+x_2(t)\right)
    \\q_2(t)=\sqrt{\frac{m}{4}}\left(x_1(t)-x_2(t)\right)\,,
    \label{normcoord}
\end{split}
\end{equation}
and with the natural (undamped) frequencies of the modes $\omega_{01}=\omega_0$ and $\omega_{02}=\sqrt{3}\omega_0$, where $\omega_0=\sqrt{k/m}$. Normalisation factors $\sqrt{m/4}$ in \eqref{normcoord} ensure that our expression \eqref{energy2D} for the energy of the system corresponds to energy expressed over the displacements and velocities of the masses, i.e. 
\begin{equation}
E(t)=\sum_{i=1}^2\left(\dot q_i(t)^2+\omega_{0i}^2q_i(t)^2\right)=\sum_{i=1}^2\left(\frac{m\dot{x}_i(t)^2}{2}+\frac{kx_i(t)^2}{2}\right)+\frac{k(x_1(t)-x_2(t))^2}{2}\,.
\label{Ephy2D}
\end{equation}

Using the specific values of undamped modal frequencies of this system, relations \eqref{gammaopt}, \eqref{avgammaopt} and \eqref{gammaopt2} become
\begin{equation}
\gamma_{\opt}(\psi,\theta_1,\theta_2)=\sqrt{\frac{3}{6\cos^2\psi\cos^2\theta_1+2\sin^2\psi\cos^2\theta_2}}\omega_0\,,
\label{gammaopt3}
\end{equation}
\begin{equation}
\overline{\gamma}_{\opt}=\frac{\sqrt{6}}{2}\omega_0\,,
\label{avgammaopt2}
\end{equation}
\begin{equation}
\gamma_{\opt}(\beta_1,\beta_2)=\sqrt{\frac{3}{6\beta_1+2\beta_2}}\omega_0\,.
\label{gammaopt4}
\end{equation}
Since $\omega_{01}=\omega_0$, the range of \eqref{gammaopt4} is $\gamma_{\opt}\in[\sqrt{2}\omega_0/2,+\infty)$ (see \eqref{rangegamma}).

\begin{figure}[h!t!]
\begin{center}
\includegraphics[width=0.48\textwidth]{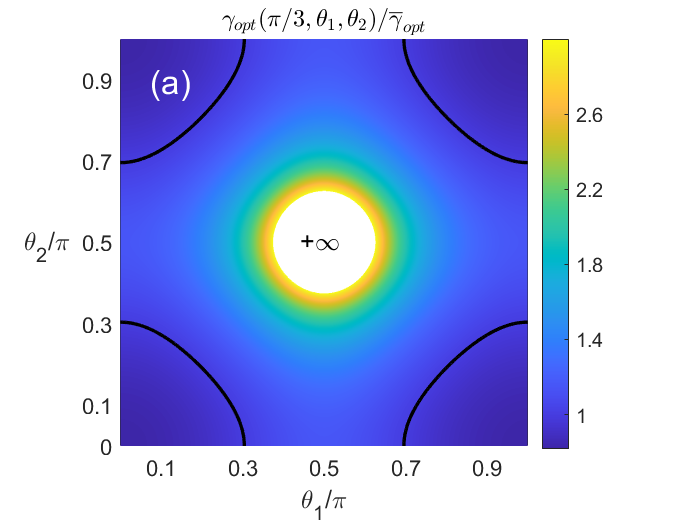}
\includegraphics[width=0.48\textwidth]{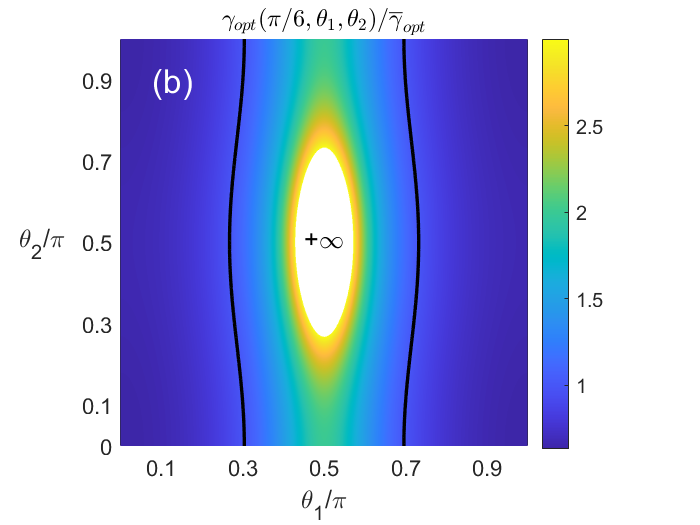}
\end{center}
\caption{Ratio $\gamma_{\opt}(\psi,\theta_1,\theta_2)/\overline{\gamma}_{\opt}$ of the optimal damping coefficients \eqref{gammaopt3} and \eqref{avgammaopt2} for $\psi=\lbrace\pi/3,\pi/6\rbrace$ and $\theta_i\in[0,\pi]$. (a) For $\psi=\pi/3$, the total initial energy $E_0$ is distributed within the modes as $E_{01}=E_0/4$ and $E_{02}=3E_0/4$. (b) For $\psi=\pi/6$, the total initial energy is distributed within the modes as $E_{01}=3E_0/4$ and $E_{02}=E_0/4$. Singularities for $\theta_1=\theta_2=\pi/2$ are indicated by infinity symbols, and the points around singularities for which $\gamma_{\opt}(\psi,\theta_1,\theta_2)/\overline{\gamma}_{\opt}>3$ are removed on both figures (central white areas). Black lines, on both figures, indicate the points for which $\gamma_{\opt}(\psi,\theta_1,\theta_2)/\overline{\gamma}_{\opt}=1$. On both figures, ratio attains minimum for the corner points, i.e. for $(\theta_1,\theta_2)=\lbrace(0,0),(0,\pi),(\pi,0),(\pi,\pi)\rbrace$.  }
\label{fig:2Dpsi}
\end{figure}

\begin{figure}[h!t!]
\begin{center}
\includegraphics[width=0.55\textwidth]{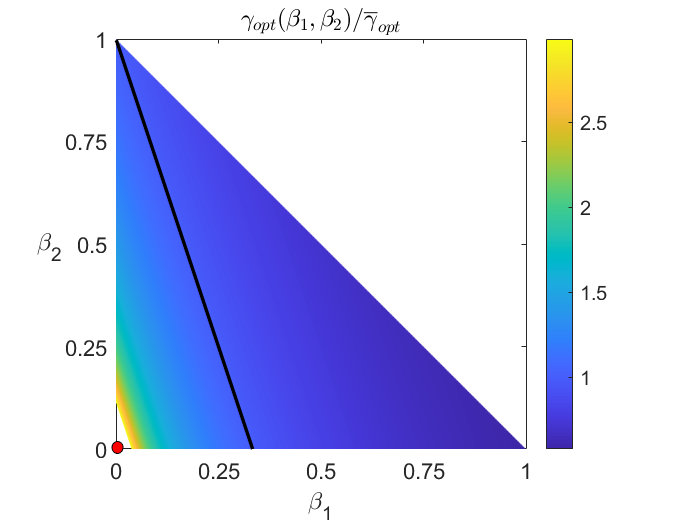}
\end{center}
\caption{Ratio $\gamma_{\opt}(\beta_1,\beta_2)/\overline{\gamma}_{\opt}$ of the optimal damping coefficients \eqref{gammaopt4} and \eqref{avgammaopt2} for $\beta_1\in[0,1]$, $\beta_2\in[0,1]$ and the constraint $0\leq(\beta_1+\beta_2)\leq 1$. Singularity for $\beta_1=\beta_2=0$ is indicated by the filled red circle, and the points near singularity, for which $\gamma_{\opt}(\beta_1,\beta_2)/\overline{\gamma}_{\opt}>3$, are removed, thus, a small white triangle is formed with the right angle at the origin. Black line indicates the points for which $\gamma_{\opt}(\beta_1,\beta_2)/\overline{\gamma}_{\opt}=1$. The minimum value of the ratio is at the point $(\beta_1,\beta_2)=(1,0)$.  }
\label{fig:2Dbeta}
\end{figure}

As examples of the behavior of the damping coefficient \eqref{gammaopt3} as a function of the angles $\lbrace\psi,\theta_1,\theta_2\rbrace$ and its relationship with the damping coefficient \eqref{avgammaopt2}, in Fig.\ \ref{fig:2Dpsi} we show $\gamma_{\opt}(\psi,\theta_1,\theta_2)/\overline{\gamma}_{\opt}$ for $\psi=\lbrace\pi/3,\pi/6\rbrace$ and $\theta_i\in[0,\pi]$. 
In Fig.\ \ref{fig:2Dbeta} we show ratio of the damping coefficient \eqref{gammaopt4} and the damping coefficient \eqref{avgammaopt2}, i.e. $\gamma_{\opt}(\beta_1,\beta_2)/\overline{\gamma}_{\opt}$.

\begin{figure}[h!t!]
\begin{center}
\includegraphics[width=0.48\textwidth]{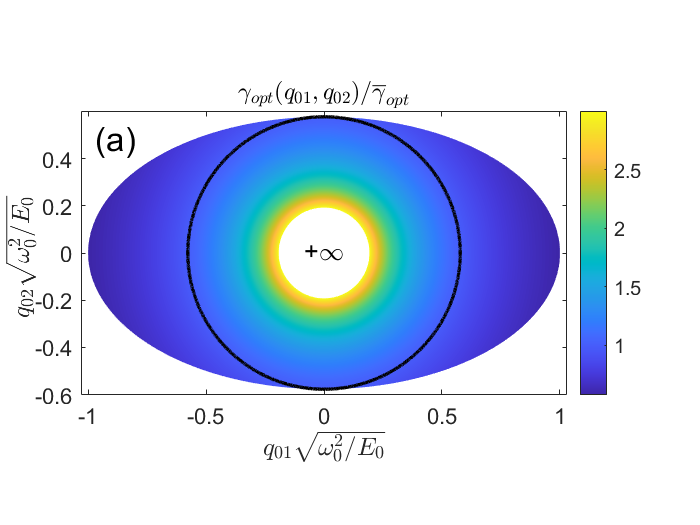}
\includegraphics[width=0.48\textwidth]{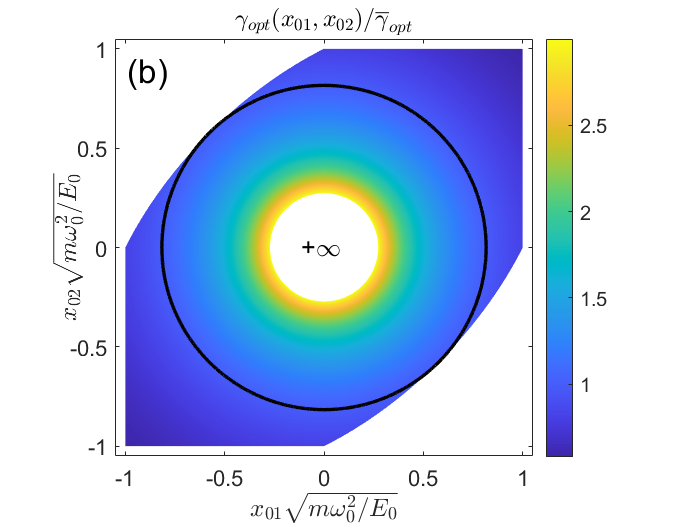}
\end{center}
\caption{(a) Ratio $\gamma_{\opt}(q_{01},q_{02})/\overline{\gamma}_{\opt}$ of the optimal damping coefficients \eqref{gammaopt_q01} and \eqref{avgammaopt2}. (b) Ratio $\gamma_{\opt}(x_{01},x_{02})/\overline{\gamma}_{\opt}$ of the optimal damping coefficients \eqref{gammaopt_X0} and \eqref{avgammaopt2}. Singularities, at points $(0,0)$ on the both figures, are denoted by infinity symbols, and the points near singularities, for which $\gamma_{\opt}/\overline{\gamma}_{\opt}>3$, are removed. Black circles on both figures indicate the points for which $\gamma_{\opt}/\overline{\gamma}_{\opt}=1$.  }
\label{fig:Q0X0}
\end{figure}

If the initial energy is comprised only of potential energy, in terms of initial modal coordinates we have $E_0=\omega_0^2q_{01}^2+3\omega_0^2q_{02}^2$, thus, the initial modal coordinates satisfy
\begin{equation}
\begin{split}
    q_{01}\sqrt{\frac{\omega_0^2}{E_0}}\in\left[-1,1\right]\,,
    \\q_{02}\sqrt{\frac{\omega_0^2}{E_0}}\in\left[-\sqrt{3}/3,\sqrt{3}/3\right]\,,
    \\0\leq \frac{\omega_0^2}{E_0}\left(q_{01}^2+3q_{02}^2\right)\leq 1\,. 
    \label{normcoordomena}
\end{split}
\end{equation}
Furthermore, we can write the optimal damping coefficient \eqref{gammaopt_q0} as    
\begin{equation}
\gamma_{\opt}(q_{01},q_{02})=\sqrt{\frac{E_0}{2\omega_0^2(q_{01}^2+q_{02}^2)}}\omega_0\,,
\label{gammaopt_q01}
\end{equation}
and the condition \eqref{conditionQ} as
\begin{equation}
    \frac{\omega_0^2}{E_0}\left(q_{01}^2+q_{02}^2\right)=\frac{1}{3}.
    \label{conditionQ1}
\end{equation}
In Fig.\ \ref{fig:Q0X0}(a) we show the ratio of \eqref{gammaopt_q01} and \eqref{avgammaopt2}, i.e. $\gamma_{\opt}(q_{01},q_{02})/\overline{\gamma}_{\opt}$. The domain of this function consists of points inside and on the ellipse, i.e. it is given by \eqref{normcoordomena}. Similarly as before, singularity at $(q_{01},q_{02})=(0,0)$ is indicated by the infinity symbol, and the points for which $\gamma_{\opt}(q_{01},q_{02})/\overline{\gamma}_{\opt}>3$ are removed. 
For points inside the circle we have $\gamma_{\opt}(q_{01},q_{02})/\overline{\gamma}_{\opt}>1$, and for points outside the circle we have $\gamma_{\opt}(q_{01},q_{02})/\overline{\gamma}_{\opt}<1$. Minimum values of this ratio are $\sqrt{3}/3\approx 0.58$, obtained for the points $(q_{01},q_{02})=\lbrace(-\sqrt{E_0}/\omega_0,0),(\sqrt{E_0}/\omega_0,0)\rbrace$. 

Using \eqref{normcoord} we can write the optimal damping coefficient \eqref{gammaopt_q01} in terms of initial displacements $x_i(0)\equiv x_{0i}$ as 
\begin{equation}
\gamma_{\opt}(x_{01},x_{02})=\sqrt{\frac{E_0}{m(x_{01}^2+x_{02}^2)}}=\sqrt{\frac{E_0}{m\omega_0^2(x_{01}^2+x_{02}^2)}}\omega_0\,.
\label{gammaopt_X0}
\end{equation}
If the initial energy is comprised only of potential energy, in terms of initial displacements we have $E_0=m\omega_0^2(x_{01}^2+x_{02}^2-x_{01}x_{02})$, thus, the initial displacements of the masses satisfy 
\begin{equation}
\begin{split}
    x_{0i}\sqrt{\frac{m\omega_0^2}{E_0}}\in\left[-1,1\right]\,,
    \\0\leq \frac{m\omega_0^2}{E_0}\left(x_{01}^2+x_{02}^2-x_{01}x_{02}\right)\leq 1\,, 
    \label{X0domena}
\end{split}
\end{equation}
and the condition \eqref{conditionQ1} is now
\begin{equation}
    \frac{m\omega_0^2}{E_0}\left(x_{01}^2+x_{02}^2\right)=\frac{2}{3}.
    \label{conditionX0}
\end{equation}
In Fig.\ \ref{fig:Q0X0}(b) we show the ratio of \eqref{gammaopt_X0} and \eqref{avgammaopt2}, i.e. $\gamma_{\opt}(x_{01},x_{02})/\overline{\gamma}_{\opt}$. The domain of this function consists of points given by \eqref{X0domena}. Similarly as before, singularity at $(x_{01},x_{02})=(0,0)$ is indicated by the infinity symbol, and the points for which $\gamma_{\opt}(x_{01},x_{02})/\overline{\gamma}_{\opt}>3$ are removed. 
For points inside the circle $\gamma_{\opt}(x_{01},x_{02})/\overline{\gamma}_{\opt}>1$, and for points outside the circle $\gamma_{\opt}(x_{01},x_{02})/\overline{\gamma}_{\opt}<1$. Minimum values of this ratio are $\sqrt{3}/3\approx 0.58$, obtained for the points $(x_{01},x_{02})=\left(\pm\sqrt{\frac{E_0}{m\omega_0^2}},\pm\sqrt{\frac{E_0}{m\omega_0^2}}\right)$.

\section{MDOF systems with MPD}

\begin{figure}[h!t!]
\begin{center}
\includegraphics[width=1\textwidth]{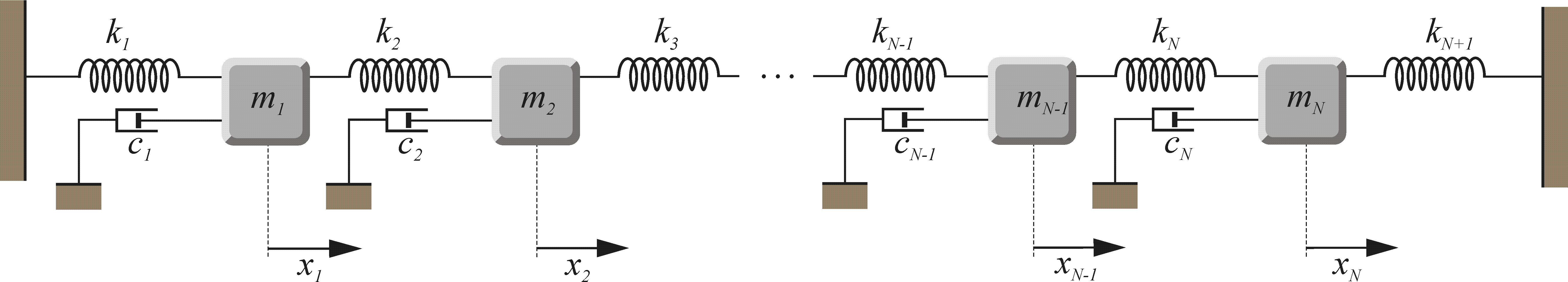}
\end{center}
\caption{Schematic figure of a MDOF system with $N$ degrees of freedom.  }
\label{fig:skicaND}
\end{figure}

Here we consider the MDOF system with $N$ degrees of freedom shown schematically in Fig.\ \ref{fig:skicaND}. As in the Section \ref{2Dsection}, we will consider MPD, i.e. masses $\lbrace m_1,m_2,...,m_N\rbrace$, spring constants $\lbrace k_1,k_2,...,k_{N+1}\rbrace$, and dampers $\lbrace c_1,c_2,...,c_N\rbrace$ can in general be mutually different but the condition $c_i/m_i=2\gamma$ holds for any $i=\lbrace1,...,N\rbrace$, where $\gamma$ is the damping coefficient. Therefore, the reasoning we presented in Section \ref{2Dsection} can be applied here, with the main difference that now the system has $N$ modes instead of two. Again, we can write each initial condition over polar coordinates, as in the 2-DOF case (see \eqref{polar2D}), only now we have $N$ pairs of polar coordinates instead of two. 
%
%

The energy of each mode is given by \eqref{Eipolar}, and consequently, the integral of the total energy over the entire time, for some arbitrary initial condition, is simply calculated similarly as in \eqref{int2D1}, i.e.  
\begin{equation}
I(\gamma, \lbrace E_{0i}\rbrace, \lbrace \theta_i\rbrace)=\sum_{i=1}^N\int_0^{\infty}E_i(t)dt=\sum_{i=1}^N\frac{E_{0i}}{2\omega_{0i}}\left(\frac{\omega_{0i}^2+\gamma^2}{\gamma\omega_{0i}}+\frac{\gamma}{\omega_{0i}}\cos2\theta_i+\sin2\theta_i\right)\,,
\label{intND1}
\end{equation}
where, again, $E_i(t)$ is the energy of the $i$-th mode, $E_{0i}$ is the initial energy of the $i$-th mode. Thus, each initial condition with energy $E_0=\sum_{i=1}^NE_{0i}$ is represented by points on $N$ circles with radii $r_i=\sqrt{E_{0i}}$, for which condition $\sum_{i=1}^Nr_i^2=E_0$ holds, and with angles $\theta_i$ that tell us how initial potential and initial kinetic energy are distributed within the modes. 

Similarly as before, initial energy of the $i$-th mode can be written as $E_{0i}=a_i^2E_0$, where coefficient $a_i^2\in[0,1]$ denotes the fraction of the initial energy of the $i$-th mode in the total initial energy $E_0$, and the condition
\begin{equation}
\sum_{i=1}^N a_i^2=1
\label{aiND}
\end{equation}
holds. Relation \eqref{aiND} defines a sphere embedded in $N$-dimensional space and we can express the coefficients $a_i$ over $N$-dimensional spherical coordinates ($N-1$ independent coordinates, i.e. angles, since the radius is equal to one), but for the sake of simplicity we will not do that here and we will stick to writing the expressions as a functions of the coefficients $a_i$. Thus, we can write \eqref{intND1} as
\begin{equation}
I(\gamma, \lbrace a_i\rbrace, \lbrace \theta_i\rbrace)=\sum_{i=1}^N\int_0^{\infty}E_i(t)dt=E_0\sum_{i=1}^N\frac{a_i^2}{2\omega_{0i}}\left(\frac{\omega_{0i}^2+\gamma^2}{\gamma\omega_{0i}}+\frac{\gamma}{\omega_{0i}}\cos2\theta_i+\sin2\theta_i\right)\,.
\label{intND2}
\end{equation}
We differentiate relation \eqref{intND2} by $\gamma$ and equate it to zero and get
\begin{equation}
\gamma_{\opt}(\lbrace a_i\rbrace,\lbrace\theta_i\rbrace)=\left(\sum_{i=1}^N\frac{2a_i^2\cos^2\theta_i}{\omega_{0i}^2}\right)^{-1/2}\,
\label{gammaND}
\end{equation}
as the optimal damping coefficient for which integral \eqref{intND2} is minimal. For any fixed set of coefficients $\lbrace a_i\rbrace$, the smallest magnitude of the function \eqref{gammaND} is obtained for $\cos^2\theta_i=1$ $\forall i$, which corresponds to the initial conditions with initial energy comprised only of potential energy distributed within the modes, i.e $E_0=\sum_{i=1}^NE_{0Pi}$. In that case the denominator of \eqref{gammaND} is
\begin{equation}
f(\lbrace a_i\rbrace)=\left(\sum_{i=1}^N\frac{2a_i^2}{\omega_{0i}^2}\right)^{1/2}\,
\label{denominatorND}
\end{equation}
and using $a_1^2=1-\sum_{i=2}^Na_i^2$ (see \eqref{aiND}) we can write \eqref{denominatorND} as
\begin{equation}
f(\lbrace a_i\rbrace)=\left(\frac{2}{\omega_{01}^2}+\sum_{i=2}^N2a_i^2\left(\frac{1}{\omega_{0i}^2}-\frac{1}{\omega_{01}^2}\right)\right)^{1/2}\,.
\label{denominatorND2}
\end{equation}
Since $\omega_{01}<\omega_{0i}$ for any $i\geq 2$, each term in the sum of relation \eqref{denominatorND2} is negative, and we can conclude that the function \eqref{denominatorND2} has maximum for the set $\lbrace a_i\rbrace=\lbrace 1,0,...,0\rbrace$. Thus, the minimum value of the optimal damping coefficient \eqref{gammaND} is $\sqrt{2}\omega_{01}/2$, and it is obtained for $a_1=1$ and $\theta_1=\lbrace0,\pi\rbrace$, which corresponds to the initial conditions with initial energy comprised only of potential energy in the first mode, i.e. $E_0=E_{0P1}$. On the other hand, for any fixed set $\lbrace a_i\rbrace$, the function \eqref{gammaND} has singularities for $\cos^2\theta_i=0$ $\forall i$. Thus, the range of the optimal damping coefficient \eqref{gammaND} is
\begin{equation}
    \gamma_{\opt}\in\left[\sqrt{2}\omega_{01}/2,+\infty\right)\,.
    \label{rangegammaN}
\end{equation}
In \ref{Appendix1} we have calculated the average of the integral \eqref{intND2} over a set of all initial conditions and obtained
\begin{equation}
\overline{I}(\gamma)=\frac{E_0}{2N}\sum_{i=1}^N\left(\frac{\omega_{0i}^2+\gamma^2}{\gamma\omega_{0i}^2}\right)\,.
\label{intNDav}
\end{equation}
We differentiate relation \eqref{intNDav} by $\gamma$ and equate it to zero and obtain
\begin{equation}
\overline{\gamma}_{\opt}=N^{1/2}\left(\sum_{i=1}^N\frac{1}{\omega_{0i}^2}\right)^{-1/2}\, 
\label{avgammaNDopt}
\end{equation}
as the optimal damping coefficient with respect to the averaged integral \eqref{intNDav}.

Since the ratio of the initial potential energy of the $i$-th mode and the total initial energy is
\begin{equation}
\beta_i=\frac{E_{0Pi}}{E_0}=a_i^2\cos^2\theta_i\,,
\label{betaND}
\end{equation}
where $\beta_i\in[0,1]$ and the condition $0\leq\sum_{i=1}^N\beta_i\leq 1$ holds, we can write \eqref{gammaND} as a function of the distribution of the initial potential energy over the modes, i.e. 
\begin{equation}
\gamma_{\opt}(\lbrace \beta_i\rbrace)=\left(\sum_{i=1}^N\frac{2\beta_i}{\omega_{0i}^2}\right)^{-1/2}\,.
\label{gammaNDpot}
\end{equation}
The minimum value of \eqref{gammaNDpot} is $\sqrt{2}\omega_{01}/2$, obtained for $\beta_1=1$ and $\beta_i=0$ for $i\geq2$, while $\gamma_{\opt}\rightarrow+\infty$ for $\beta_i=0$ $\forall i$. Using $\beta_i=\omega_{0i}^2q_{0i}^2/E_0$, we can write  \eqref{gammaND} as a function of initial modal coordinates as well, i.e.
\begin{equation}
\gamma_{\opt}(\lbrace q_{0i}\rbrace)=\sqrt{\frac{E_0}{2\sum_{i=1}^Nq_{0i}^2}}\,,
\label{gammaNDq}
\end{equation}
where $q_{0i}\in[-\sqrt{E_0}/\omega_{0i},\sqrt{E_0}/\omega_{0i}]$ and the condition $0\leq\sum_{i=1}^N\omega_{0i}^2q_{0i}^2\leq E_0$ holds.

\subsection{Quantitative example}
\label{exampleND}

\begin{figure}[h!t!]
\begin{center}
\includegraphics[width=0.48\textwidth]{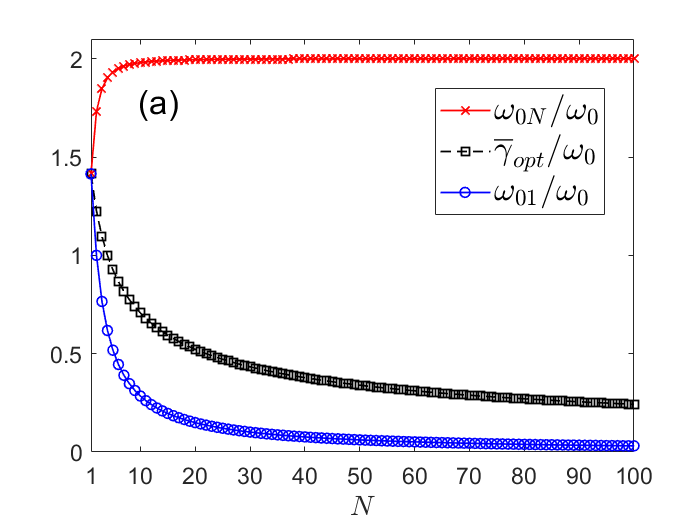}
\includegraphics[width=0.48\textwidth]{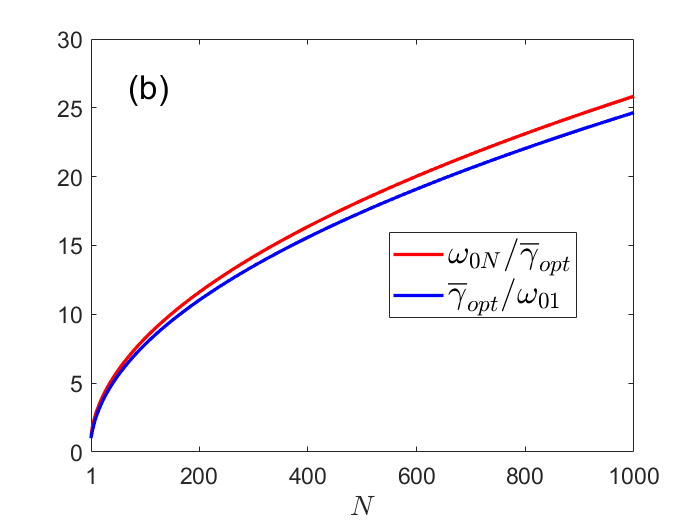}
\end{center}
\caption{(a) Undamped modal frequencies $\omega_{01}$ (blue circles), $\omega_{0N}$ (red x's) and the damping coefficient $\overline{\gamma}_{\opt}$ (black squares) as functions of the number of the masses $N$. (b) Ratios $\overline{\gamma}_{\opt}/\omega_{01}$ (blue line) and $\omega_{0N}/\overline{\gamma}_{\opt}$ (red line), shown as solid lines due to the high density of the shown points.  }
\label{fig:NvsMod1}
\end{figure}

Here we consider the MDOF system as the one shown schematically in Fig.\ \ref{fig:skicaND} but with $m_i=m$, $c_i=c$ for $i=\lbrace1,...,N\rbrace$, and with $k_i=k$ for $i=\lbrace1,...,N+1\rbrace$. Such a system without damping, i.e with $c_i=0$ $\forall i$, is a standard part of the undergraduate physics/mechanics courses \cite{Berkeley}. Therefore, for the MDOF system with $N$ degrees of freedom we are considering here, the undamped modal frequencies are \cite{grigoriu2021linear, Berkeley}
\begin{equation}
    \omega_{0i}=2\omega_0\sin\left(\frac{i\pi}{2(N+1)}\right)\,,\textrm{with}\,\,i=\lbrace1,...,N\rbrace\,,
    \label{modoviN}
\end{equation}
and where $\omega_0=\sqrt{k/m}$. In Fig.\ \ref{fig:NvsMod1}(a) we show undamped modal frequencies $\omega_{01}$, $\omega_{0N}$ and damping coefficient $\overline{\gamma}_{\opt}$, i.e. \eqref{avgammaNDopt}, calculated with \eqref{modoviN}, as functions of $N$. We clearly see that the coefficient $\overline{\gamma}_{\opt}$ is in the over-damped regime from the perspective of the first mode, and in the under-damped regime from the perspective of highest mode, for any $N>1$, and in the case $N=1$ all three values match. In Fig.\ \ref{fig:NvsMod1}(b) we show ratios $\overline{\gamma}_{\opt}/\omega_{01}$ and $\omega_{0N}/\overline{\gamma}_{\opt}$ and we see that both ratios increase with increasing $N$.

We show in \ref{Appendix2} that the following limits hold
\begin{equation}
    \lim_{N \to +\infty} \overline{\gamma}_{\opt}(N)=0,
    \label{limes1}
\end{equation}
\begin{equation}
    \lim_{N \to +\infty} \frac{\overline{\gamma}_{\opt}(N)}{\omega_{01}(N)}=+\infty,
    \label{limes2}
\end{equation}
\begin{equation}
    \lim_{N \to +\infty} \frac{\omega_{0N}(N)}{\overline{\gamma}_{\opt}(N)}=+\infty.
    \label{limes3}
\end{equation}
We note here that these limit values do not correspond to the transition from a discrete to a continuous system, but simply tell us the behavior of these quantities with respect to the increase in the number of masses, i.e. with respect to the increase in the size of the discrete system.

From everything that has been said so far, it is clear that the damping coefficient $\overline{\gamma}_{\opt}$, obtained by minimizing the energy integral averaged over all initial conditions that correspond to the same initial energy, cannot be considered generally as optimal and that, by itself, it says nothing about optimal damping of the system whose dynamics started with some specific initial condition. 
Damping coefficient \eqref{gammaND}, which is given by the minimization of the energy integral for a specific initial condition, is of course a better choice for optimal damping of an MDOF system, than the damping coefficient $\overline{\gamma}_{\opt}$, if we want to consider how the system dissipates energy the fastest for a particular initial condition, but, as we argue in the subsection \ref{issues}, this damping coefficient also has some obvious deficiencies.

\subsection{Issues with the minimum of the energy integral as a criterion for optimal damping}
\label{issues}

We can ask, for example, whether in an experiment, with known initial conditions, in which an MDOF system is excited to oscillate, a damping coefficient \eqref{gammaND} would be the best choice if we want that the system settles down in equilibrium as soon as possible? Here, in three points, we explain why we think the answer to that question is negative:
\begin{itemize}
\item From relation \eqref{intND2}, we see that, due to the term $\sin2\theta_i$, the energy integral is sensitive to changes $\theta_i\rightarrow -\theta_i$ and $\theta_i\rightarrow\pi-\theta_i$, which correspond to changes of initial conditions $(q_{0i},\dot{q}_{0i})\rightarrow(q_{0i},-\dot{q}_{0i})$ and $(q_{0i},\dot{q}_{0i})\rightarrow(-q_{0i},\dot{q}_{0i})$. When we differentiate \eqref{intND2} to determine $\gamma$ for which the energy integral has a minimum, the term $\sin2\theta_i$ cancels and as a result the coefficient \eqref{gammaND} is not sensitive to this change in initial conditions. Such changes in the initial conditions lead to significantly different situations. For example, if $q_{0i}>0$ and $\dot{q}_{0i}>0$, the $i$-th mode in the critical and over-damped regime (i.e. for $\gamma\geq\omega_{0i}$) will never reach the equilibrium position, while for $q_{0i}>0$ and $\dot{q}_{0i}<0$, and $i$-th mode initial kinetic energy grater than initial potential energy, it can go through the equilibrium position once, depending on the magnitude of the damping coefficient, and there will be the smallest damping coefficient in the over-damped regime for which no crossing occurs and for which the solution converges to equilibrium faster than for any other damping coefficient \cite{Lelas}. Therefore, the damping coefficient considered optimal would have to be sensitive to this change in initial conditions. 
\item Damping coefficient \eqref{gammaND} has singularities for $\cos\theta_i=0$ $\forall i$, i.e. for initial conditions for which all initial energy is kinetic. For such initial conditions, the higher the damping coefficient, the higher and faster the dissipation. In other words, the higher the damping coefficient, the faster the energy integral decreases. Therefore, coefficient \eqref{gammaND} diverges for that type of initial conditions. This would actually mean that, for this initial conditions, it is optimal to take the damping coefficient as high as possible, but in principle this corresponds to a situation in which all modes are highly over-damped, i.e. all masses reach their maximum displacements in a very short time and afterwards they begin to return to the equilibrium position almost infinitely slowly. Figuratively speaking, it is as if we immersed the system in concrete. This issue has recently been addressed in the context of free vibrations of SDOF \cite{Lelas} and was already noticed in \cite{nakic2019mixed}. Therefore, simply taking the highest possible damping coefficient, as suggested by relation \eqref{gammaND} for this type of initial conditions, is not a good option.
\item The damping coefficient \eqref{gammaND} is determined on the basis of the energy integral over the entire time and therefore it does not take into account that in nature and experiments these systems effectively return to the equilibrium state for some finite time.
\end{itemize}

Because of the above points, in the next section we provide a new approach to determine the optimal damping of MDOF systems.

\section{Optimal damping of an MDOF system: a new perspective}
\label{newopt}

From a theoretical perspective, systems with viscous damping asymptotically approach the equilibrium state and never reach it exactly. In nature and in experiments, these systems reach the equilibrium state which is not an exact zero energy state, but rather a state in which the energy of the system has decreased to the level of the energy imparted to the system by the surrounding noise, or to the energy resolution of the measuring apparatus. Following this line of thought, we will define a system to be in equilibrium for times $t>\tau$ such that
\begin{equation}
\frac{E(\tau)}{E_0}=10^{-\delta}\,,
\label{conditiontau}
\end{equation}
where $E(\tau)$ is the energy of the system at $t=\tau$, $E_0$ is the initial energy, and $\delta>0$ is a dimensionless parameter that defines what fraction of the initial energy is left in the system. This line of thought has recently been used to determine the optimal damping of SDOF systems \cite{Lelas}, and here we extend it to MDOF systems. Therefore, in what follows, we will consider as optimal the damping coefficient for which the systems energy drops to some energy level of interest, e.g. to the energy resolution of the experiment, the fastest and we will denote it with $\tilde{\gamma}$.

\subsection{Optimal damping of the $i$-th mode of a MDOF system with MPD}
\label{optimal1D}

Here we will consider the behavior of the energy of the $i$-th mode of the MDOF system  with MPD and determine the optimal damping coefficient $\tilde{\gamma}_i$ of the $i$-th mode with respect to criterion \eqref{conditiontau}. For any MDOF system with $N\geq1$ degrees of freedom with MPD, each mode behaves as a SDOF system studied in Section \ref{1Dsection}, with the damping coefficient $\gamma$ and the undamped (natural) frequency $\omega_{0i}$. Thus (see relation \eqref{Eipolar}), the ratio of the energy of the $i$-th mode, $E_i(\gamma,t)$, and initial energy of the $i$-th mode, $E_{0i}$, is given by
\begin{equation}
\frac{E_i(\gamma, t)}{E_{0i}}=e^{-2\gamma t}\left( \cos^2(\omega_i t)+\gamma\cos2\theta_i\frac{\sin(2\omega_i t)}{\omega_i}+\left(\omega_{0i}^2+\gamma^2+2\omega_{0i}\gamma\sin2\theta_i\right)\frac{\sin^2(\omega_i t)}{\omega_i^2}\right)\,
\label{Eit2}
\end{equation}
for the under-damped ($\gamma<\omega_{oi}$) and over-damped ($\gamma>\omega_{0i}$) regime. We will repeat here briefly for clarity, $\omega_i=\sqrt{\omega_{0i}^2-\gamma^2}$ is the damped angular frequency and $\theta_i$ is the polar angle which determines the initial conditions $q_{0i}$ and $\dot{q}_{0i}$ of the $i$-th mode and the distribution of the initial energy within the mode, i.e. initial potential and initial kinetic energy of the $i$-th mode are $E_{0Pi}=E_{0i}\cos^2\theta_i$ and $E_{0Ki}=E_{0i}\sin^2\theta_i$ respectively. Energy to initial energy ratio for the $i$-th mode in the critically damped regime ($\gamma=\omega_{0i}$) is simply obtained by taking $\gamma\rightarrow\omega_{0i}$ limit of the relation \eqref{Eit2}, and we obtain
\begin{equation}
\frac{E_i(\gamma=\omega_{0i}, t)}{E_{0i}}=e^{-2\omega_{0i} t}\left( 1+2\omega_{0i}(\cos2\theta_i)t+2\omega_{0i}^2\left(1+\sin2\theta_i\right)t^2\right)\,.
\label{Eit2c}
\end{equation}
In relations \eqref{Eit2} and \eqref{Eit2c}, we explicitly show that the energy depends on the damping coefficient and time, because in what follows we will plot these quantities as functions of these two variables for fixed initial conditions, i.e. fixed $\theta_i$. We will investigate the behavior for several types of initial conditions, which of course will not cover all possible types of initial conditions, but will give us a sufficiently clear picture of the determination and behavior of the optimal damping with respect to the initial conditions and the equilibrium state defined with condition \eqref{conditiontau}.

\subsubsection{Initial energy of the $i$-th mode comprised only of potential energy}
\label{subsub1}

\begin{figure}[h!t!]
\begin{center}
\includegraphics[width=0.55\textwidth]{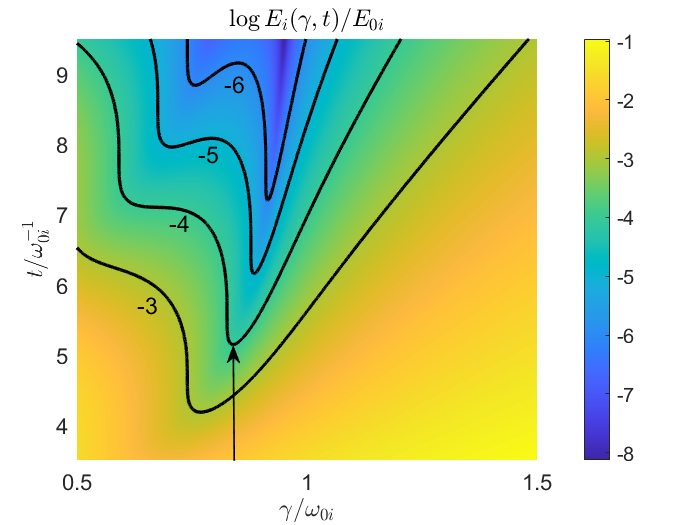}
\end{center}
\caption{The base $10$ logarithm of the ratio \eqref{Eit2}, i.e. $\log\left(E_i(\gamma, t)/E_{0i}\right)$, for initial condition $\theta_i=0$. For this initial condition, initial energy of the $i$-th mode is comprised only of potential energy. Four black contour lines denote points with $E_i(\gamma,t)/E_{0i}=\lbrace 10^{-3},10^{-4},10^{-5},10^{-6}\rbrace$ respectively, as indicated by the numbers placed to the left of each contour line. As an example of determining the optimal damping for which the system reaches the desired energy level the fastest,  i.e. with respect to the condition \eqref{conditiontau}, we draw the arrow that points to the coordinates $(\gamma,t)=(0.840\omega_{0i},5.15\omega_{0i}^{-1})$ for which the $i$-th mode reaches the level $E_i(\gamma,t)/E_{0i}=10^{-4}$ the fastest. Thus, $\tilde{\gamma}_i=0.840\omega_{0i}$ is the optimal damping coefficient to reach this energy level the fastest. Optimal values for other energy levels, denoted with contour lines, are given in Table \ref{table1D1}.  }
\label{fig:EPmin1D}
\end{figure}

In Fig.\ \ref{fig:EPmin1D} we show the base $10$ logarithm of the ratio \eqref{Eit2}, i.e. $\log\left(E_i(\gamma, t)/E_{0i}\right)$, for initial condition $\theta_i=0$, which corresponds to the initial energy of the $i$-th mode comprised only of potential energy. Four black contour lines denote points with $E_i(\gamma,t)/E_{0i}=\lbrace10^{-3},10^{-4},10^{-5},10^{-6}\rbrace$ respectively, as indicated by the numbers placed to the left of each contour line. Each contour line has a unique point closest to the $\gamma$ axis, i.e. corresponding to the damping coefficient $\tilde{\gamma}_i$ for which that energy level is reached the fastest. As an example, we draw arrow in Fig.\ \ref{fig:EPmin1D} that points to the coordinates $(\gamma,t)=(0.840\omega_{0i},5.15\omega_{0i}^{-1})$, i.e. to the tip of the contour line with points corresponding to $E_i(\gamma, t)=10^{-4}E_{0i}$. 
Thus, for the initial condition $\theta_i=0$, $\tilde{\gamma}_i=0.840\omega_{0i}$ is the optimal damping coefficient for the $i$-th mode to reach this energy level the fastest, and it does so at the instant $\tau_i=5.15\omega_{0i}^{-1}$. In Table \ref{table1D1} we show results for other energy levels corresponding to contour lines shown in Fig.\ \ref{fig:EPmin1D}. Here, and in the rest of the paper, we have rounded the results for the damping coefficient to three decimal places, and for the time to two decimal places.
%
\begingroup
\setlength{\tabcolsep}{5.2pt} 
\renewcommand{\arraystretch}{1.5} 
\begin{table}[h!t!]
\centering
\begin{tabular}{|c|c|c|} 
 \hline
 $E_i(\gamma,t)/E_{0i}$ & $\tilde{\gamma}_i \left[ \omega_{0i} \right]$ & $\tau_i [ \omega_{0i}^{-1}]$ \\ 
 \thickhline
 $10^{-3}$ & \ $0.769$ \ & $4.18$ \\ 
 \hline
 $10^{-4}$ & \ $0.840$ \ & $5.15$ \\ 
 \hline
 $10^{-5}$ & \ $0.885$ \ & $6.16$ \\ 
 \hline
 $10^{-6}$ & \ $0.915$ \ & $7.20$ \\ 
 \hline
\end{tabular}
\caption{Optimal damping coefficient $\tilde{\gamma}_i$ for which the energy of the $i$-th mode drops to the level $10^{-\delta}E_{0i}$ the fastest, with the initial condition $\theta_i=0$. 
}
\label{table1D1}
\end{table}
\endgroup

Consider now, for example, a thought experiment in which we excite a MDOF system so that it vibrates only in the first mode and that all initial energy was potential, i.e. $E_{01}=E_0$ and $\theta_1=0$. Furthermore, suppose that the system has effectively returned to equilibrium when its energy drops below $10^{-6}E_0$, due to the resolution of the measuring apparatus. It is clear form the Table \ref{table1D1} that $\tilde{\gamma}_1=0.915\omega_{01}$ would be optimal in such a scenario. In the same scenario, optimal damping coefficient given by the minimization of the energy integral, i.e. \eqref{gammaND}, would be $\gamma_{\opt}=\sqrt{2}\omega_{01}/2=0.707\omega_{01}$, thus, a very bad choice in the sense that this damping coefficient would not be optimal even in an experiment with a significantly poorer energy resolution (see Table \ref{table1D1}). This simple example illustrates that, from a practical point of view, one has to take into account both the initial conditions and the resolution of the measuring apparatus in order to determine the optimal damping coefficient. 

\subsubsection{Initial energy of the $i$-th mode comprised only of kinetic energy}
\label{subsub2}

\begin{figure}[h!t!]
\begin{center}
\includegraphics[width=0.48\textwidth]{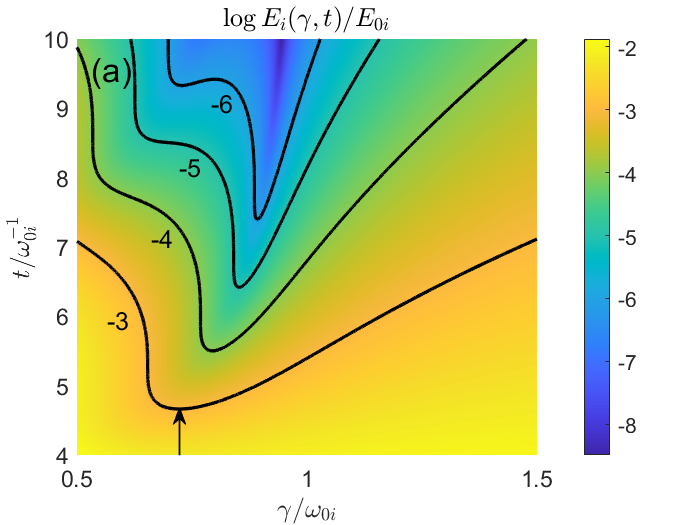}
\includegraphics[width=0.48\textwidth]{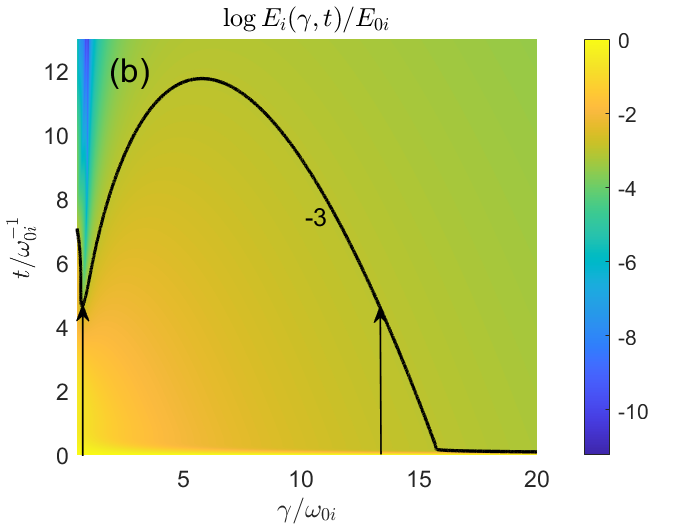}
\end{center}
\caption{The base $10$ logarithm of the ratio \eqref{Eit2}, i.e. $\log\left(E_i(\gamma, t)/E_{0i}\right)$, for initial condition $\theta_i=\pi/2$. For this initial condition, initial energy of the $i$-th mode is comprised only of kinetic energy. (a) Four black contour lines denote points with $E_i(\gamma,t)/E_{0i}=\lbrace10^{-3},10^{-4},10^{-5},10^{-6}\rbrace$ respectively, and the arrow points to the coordinates $(\gamma,t)=(0.722\omega_{0i},4.66\omega_{0i}^{-1})$, with $E_i(\gamma,t)/E_{0i}=10^{-3}$, for which this level of energy is reached in shortest time for the shown data span. (b) Contour line for points with $E_i(\gamma, t)=10^{-3}E_{0i}$ is shown for larger data span, left arrow points to the coordinates $(\gamma,t)=(0.722\omega_{0i},4.66\omega_{0i}^{-1})$, and the right arrow to the coordinates $(\gamma,t)=(13.316\omega_{0i},4.66\omega_{0i}^{-1})$, both with $E_i(\gamma,t)/E_{0i}=10^{-3}$. Thus, for $\gamma>13.316\omega_{0i}$ energy level $10^{-3}E_{0i}$ is reached faster than for $\gamma=0.7223\omega_{0i}$. See text for details.}
\label{fig:EKmin1D}
\end{figure}

In Fig.\ \ref{fig:EKmin1D}(a) and (b) we show the base $10$ logarithm of the ratio \eqref{Eit2}, i.e. $\log\left(E_i(\gamma, t)/E_{0i}\right)$, for initial condition $\theta_i=\pi/2$, which corresponds to the initial energy of the $i$-th mode comprised only of kinetic energy. 
In Fig.\ \ref{fig:EKmin1D}(b) we show results for larger data span than in Fig.\ \ref{fig:EKmin1D}(a), and only contour line for points corresponding to $E_i(\gamma, t)=10^{-3}E_{0i}$. The left arrow in Fig.\ \ref{fig:EKmin1D}(b) indicates the same coordinates as the arrow in Fig.\ \ref{fig:EKmin1D}(a), and the right arrow in Fig.\ \ref{fig:EKmin1D}(b) points to the coordinates $(\gamma,t)=(13.316\omega_{0i},4.66\omega_{0i}^{-1})$ with $E_i(\gamma, t)=10^{-3}E_{0i}$. Thus, for $\gamma>13.316\omega_{0i}$ the system comes sooner to the energy level $10^{-3}E_{0i}$ than for $\gamma=0.722\omega_{0i}$, but these highly over-damped damping coefficients would correspond to restricting the system to infinitesimal displacements from equilibrium, after which the system returns to the equilibrium state practically infinitely slowly \cite{Lelas}. Thus, for this initial condition we take the damping coefficient in the under-damped regime, i.e. $\tilde{\gamma}_i=0.722\omega_{0i}$, as optimal for reaching the level $E_i(\gamma, t)=10^{-3}E_{0i}$ the fastest. 
For all energy levels the behaviour is qualitatively the same, and the results are given in Table \ref{table1D2}.   
\begingroup
\setlength{\tabcolsep}{5.2pt} 
\renewcommand{\arraystretch}{1.5} 
\begin{table}[h!t!]
\centering
\begin{tabular}{|c|c|c|} 
 \hline
 $E_i(\gamma,t)/E_{0i}$ & $\tilde{\gamma}_i \left[ \omega_{0i} \right]$ & $\tau_i [ \omega_{0i}^{-1}]$ \\ 
 \thickhline
 $10^{-3}$ & \ $0.722$ \ & $4.66$ \\ 
 \hline
 $10^{-4}$ & \ $0.794$ \ & $5.50$ \\ 
 \hline
 $10^{-5}$ & \ $0.852$ \ & $6.42$ \\ 
 \hline
 $10^{-6}$ & \ $0.892$ \ & $7.40$ \\ 
 \hline
\end{tabular}
\caption{Optimal damping coefficient $\tilde{\gamma}_i$ for which the energy of the $i$-th mode drops to the level $10^{-\delta}E_{0i}$ the fastest, with the initial condition $\theta_i=\pi/2$. 
} 
\label{table1D2}
\end{table}
\endgroup

Consider now, for example, a thought experiment in which we excite a MDOF system so that it vibrates only in the first mode and that all initial energy was kinetic, i.e. $E_{01}=E_0$ and $\theta_1=\pi/2$. Furthermore, suppose that the system has effectively returned to equilibrium when its energy drops below $10^{-6}E_0$, due to the resolution of the measuring apparatus. It is clear form the Table \ref{table1D2} that $\tilde{\gamma}_1=0.892\omega_{01}$ would be optimal in such a scenario. In the same scenario, optimal damping coefficient given by the minimization of the energy integral, i.e. \eqref{gammaND}, would be $\gamma_{\opt}=+\infty$.

Here we note that if in such an experiment we can set the damping coefficient to be in the over-damped regime in the first part of the motion, i.e. when the system is moving from the equilibrium position to the maximum displacement, and in the under-damped regime in the second part of the motion, i.e. when the system moves from the position of maximum displacement back towards the equilibrium position, then the fastest way to achieve equilibrium would be to take the largest experimentally available over-damped coefficient in the first part of the motion, and the under-damped coefficient optimised like in \ref{subsub1} in the second part of the motion, with the fact that we have to carry out the optimization with respect to the energy left in the system at the moment when the system reached the maximum displacement and with respect to the energy resolution of the experiment.

\subsubsection{Initial energy of the $i$-th mode comprised of potential and kinetic energy}
\label{subsub3}

\begin{figure}[h!t!]
\begin{center}
\includegraphics[width=0.48\textwidth]{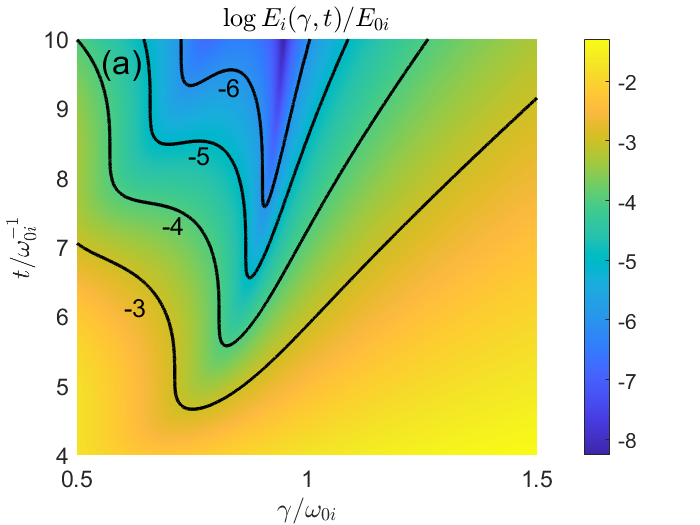}
\includegraphics[width=0.48\textwidth]{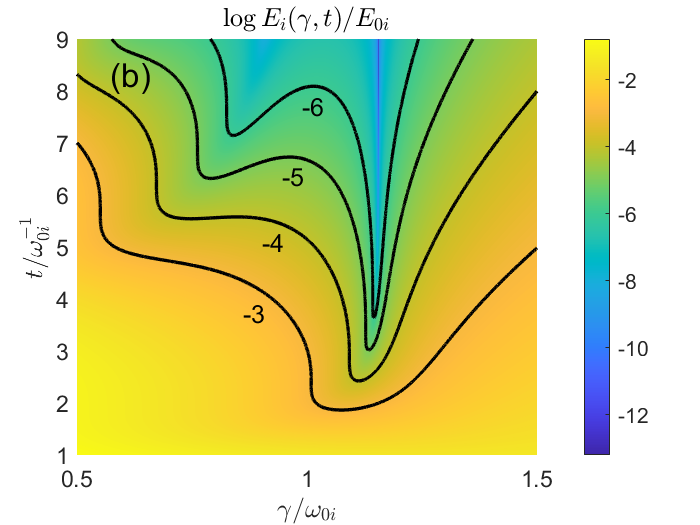}
\end{center}
\caption{The base $10$ logarithm of the ratio \eqref{Eit2}, i.e. $\log\left(E_i(\gamma, t)/E_{0i}\right)$, (a) for initial condition $\theta_i=\pi/3$, and (b) for initial condition $\theta_i=-\pi/3$. For both initial conditions, initial energy of the $i$-th mode is comprised of kinetic energy $E_{0Ki}=3E_{0i}/4$ and potential energy $E_{0Pi}=E_{0i}/4$. For $\theta_i=\pi/3$ initial normal coordinate and velocity are of the same signs, i.e. $q_{0i}>0$ and $\dot{q}_{0i}>0$. For $\theta_i=-\pi/3$ initial normal coordinate and velocity are of the opposite signs, i.e. $q_{0i}>0$ and $\dot{q}_{0i}<0$.  }
\label{fig:EKEPmin1D}
\end{figure}

In Fig.\ \ref{fig:EKEPmin1D}(a) we show the base $10$ logarithm of the ratio \eqref{Eit2}, i.e. $\log\left(E_i(\gamma, t)/E_{0i}\right)$, for initial condition $\theta_i=\pi/3$, which corresponds to the initial energy of the $i$-th mode comprised of kinetic energy $E_{0Ki}=3E_{0i}/4$ and potential energy $E_{0Pi}=E_{0i}/4$, with both initial normal coordinate and velocity positive, i.e. with $q_{0i}>0$ and $\dot{q}_{0i}>0$. 
The results for optimal damping are obtained by the same method as in \ref{subsub1} and are given in Table \ref{table3} for data shown in Fig.\ \ref{fig:EKEPmin1D}(a), and in Table \ref{table4} for data shown in Fig.\ \ref{fig:EKEPmin1D}(b). We see that the energy dissipation strongly depends on the relative sign between $q_{0i}$ and $\dot{q}_{0i}$. It was recently shown, for free vibrations of SDOF, that for an initial condition with initial kinetic energy greater than initial potential energy and opposite signs between $x_0$ and $v_0$, an optimal damping coefficient can be found in the over-damped regime \cite{Lelas}, thus, the same is true when we consider any mode of a MDOF system with MPD.   
%
%
\begin{table}[ht]
\parbox{.48\linewidth}{
\centering
\begin{tabular}{|c|c|c|} 
 \hline
 $E_i(\gamma,t)/E_{0i}$ & $\tilde{\gamma}_i \left[ \omega_{0i} \right]$ & $\tau_i [ \omega_{0i}^{-1}]$ \\ 
 \thickhline
 $10^{-3}$ & \ $0.751$ \ & $4.66$ \\ 
 \hline
 $10^{-4}$ & \ $0.825$ \ & $5.58$ \\ 
 \hline
 $10^{-5}$ & \ $0.875$ \ & $6.55$ \\ 
 \hline
 $10^{-6}$ & \ $0.908$ \ & $7.58$ \\ 
 \hline
\end{tabular}
\caption{Optimal damping coefficient $\tilde{\gamma}_i$ for which the energy of the $i$-th mode drops to the level $10^{-\delta}E_{0i}$ the fastest, with the initial condition $\theta_i=\pi/3$.\label{table3}}
} 
\hfill
\parbox{.48\linewidth}{
\centering
\begin{tabular}{|c|c|c|} 
 \hline
 $E_i(\gamma,t)/E_{0i}$ & $\tilde{\gamma}_i \left[ \omega_{0i} \right]$ & $\tau_i [ \omega_{0i}^{-1}]$ \\ 
 \thickhline
 $10^{-3}$ & \ $1.075$ \ & $1.87$ \\ 
 \hline
 $10^{-4}$ & \ $1.112$ \ & $2.42$ \\ 
 \hline
 $10^{-5}$ & \ $1.135$ \ & $3.02$ \\ 
 \hline
 $10^{-6}$ & \ $1.145$ \ & $3.64$ \\ 
 \hline
\end{tabular}
\caption{Optimal damping coefficient $\tilde{\gamma}_i$ for which the energy of the $i$-th mode drops to the level $10^{-\delta}E_{0i}$ the fastest, with the initial condition $\theta_i=-\pi/3$.\label{table4}}}
\end{table}

Consider now, for example, a thought experiment in which we excite a MDOF system so that it vibrates only in the first mode and that $75\%$ of initial energy was kinetic and $25\%$ of initial energy was potential, and with $q_{01}>0$ and $\dot{q}_{01}>0$, i.e. $E_{01}=E_0$ and $\theta_1=\pi/3$. Furthermore, suppose that the system has effectively returned to equilibrium when its energy drops below $10^{-6}E_0$, due to the resolution of the measuring apparatus. It is clear form the Table \ref{table3} that $\tilde{\gamma}_1=0.908\omega_{01}$ would be optimal in such a scenario. In the same scenario, but with $q_{01}>0$ and $\dot{q}_{01}<0$, i.e. for $\theta_1=-\pi/3$, we see from Table \ref{table4} that $\tilde{\gamma}_1=1.145\omega_{01}$ would be optimal. Optimal damping coefficient given by the minimization of the energy integral, i.e. \eqref{gammaND}, is insensitive to the change of the sign of $\dot{q}_{01}$, and it would be $\gamma_{\opt}=\sqrt{2}\omega_{01}=1.414\omega_{01}$ in both cases.

\begin{figure}[h!t!]
\begin{center}
\includegraphics[width=0.55\textwidth]{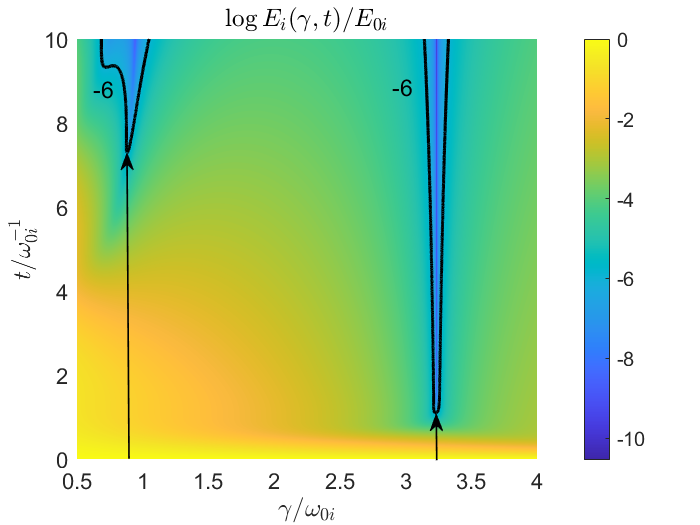}
\end{center}
\caption{The base $10$ logarithm of the ratio \eqref{Eit2}, i.e. $\log\left(E_i(\gamma, t)/E_{0i}\right)$, for initial condition $\theta_i=-9\pi/20$. Black contour line denotes the points with $E_i(\gamma,t)=10^{-6}E_{0i}$. Left arrow points to the coordinates $(\gamma,t)=(0.883\omega_{0i},7.30\omega_{0i}^{-1})$ for which level $10^{-6}E_{0i}$ is reached the fastest in the under-damped regime, and the right arrow points to the coordinates $(\gamma,t)=(3.222\omega_{0i},0.87\omega_{0i}^{-1})$ for which the same level is reached the fastest in the over-damped regime.  }
\label{fig:EK>>EP}
\end{figure}

We note here, that for the initial conditions of the $i$-th mode with initial kinetic energy much grater than initial potential energy, i.e. $E_{0Ki}>>E_{0Pi}$, and with opposite signs of initial displacement and velocity, i.e. $\mathop{\mathrm{sgn}}(q_{0i})\neq \mathop{\mathrm{sgn}}(\dot{q}_{0i})$, the optimal damping coefficient is going to be deep in the over-damped regime and dissipation of initial energy will happen in a very short time. If, for any reason, this is not desirable in some particular application, one can always find damping coefficient in the under-damped regime, with that same initial condition, which can serve as an alternative. As an example of such a situation, in Fig.\ \ref{fig:EK>>EP} we show the base 10 logarithm of the ratio \eqref{Eit2}, i.e. $\log\left(E_i(\gamma, t)/E_{0i}\right)$, for initial condition $\theta_i=-9\pi/20$, which corresponds to the initial energy of the $i$-th mode comprised of kinetic energy $E_{0Ki}\approx0.98E_{0i}$ and potential energy $E_{0Pi}\approx0.02E_{0i}$, with $q_{0i}>0$ and $\dot{q}_{0i}<0$. 
In Fig.\ \ref{fig:EK>>EP} we see that the $i$-th mode will reach the energy level $10^{-6}E_{0i}$ the fastest for $\gamma=3.222\omega_{0i}$, and in case, e.g., that such damping coefficient is difficult to achieve experimentally, another choice for the optimal damping coefficient can be $\gamma=0.883\omega_{0i}$.       

\subsection{Optimal damping of a MDOF system with MPD}
\label{optimal2D}

If all modes of a MDOF system with $N$ degrees of freedom are excited, the ratio of the energy of the system, $E(\gamma,t)$, and initial energy of the system, $E_0$, is given by
\begin{equation}
\frac{E(\gamma, t)}{E_0}=\sum_{i=1}^N\frac{E_{0i}}{E_0}e^{-2\gamma t}\left( \cos^2(\omega_i t)+\gamma\cos2\theta_i\frac{\sin(2\omega_i t)}{\omega_i}+\left(\omega_{0i}^2+\gamma^2+2\omega_{0i}\gamma\sin2\theta_i\right)\frac{\sin^2(\omega_i t)}{\omega_i^2}\right)\,,
\label{EtN}
\end{equation}
where the set of all initial energies of the modes, i.e. $\lbrace E_{0i}\rbrace$, and the set of all polar angles, i.e. $\lbrace \theta_i\rbrace$, determines the initial condition of the whole system. Since for MPD the damping of the system as a whole is determined by only one damping coefficient $\gamma$, we can calculate the base $10$ logarithm of the ratio \eqref{EtN}, but using a unique units for $\gamma$, $t$ and $\omega_{0i}$ for all modes, and from these data determine the optimal damping coefficient $\tilde{\gamma}$, for which the system will come to equilibrium in the sense of the condition \eqref{conditiontau} the fastest, in the same way as in subsubsections \ref{subsub1}-\ref{subsub3} where we showed how to determine the optimal damping of individual modes. One practical choice for the units might be $\omega_{01}$ for $\gamma$ and for $\omega_{0i}$ $\forall i$, and $\omega_{01}^{-1}$ for $t$. This way, we have the easiest insight into the relationship between the first mode and the optimal damping coefficient that we want to determine, in the sense that we can easily see whether the first mode is under-damped, over-damped or critically damped in relation to it, which is important since the first mode is often the dominant mode. If we apply this to the 2-DOF system studied in \ref{example2D}, we obtain
\begin{equation}
\frac{E(\gamma, t)}{E_0}=\sum_{i=1}^2\frac{E_{0i}}{E_0}e^{-2\gamma t}\left( \cos^2(\omega_i t)+\gamma\cos2\theta_i\frac{\sin(2\omega_i t)}{\omega_i}+\left(\omega_{0i}^2+\gamma^2+2\omega_{0i}\gamma\sin2\theta_i\right)\frac{\sin^2(\omega_i t)}{\omega_i^2}\right)\,,
\label{Et2D}
\end{equation}
where $\omega_{01}=\omega_0$, $\omega_{02}=\sqrt{3}\omega_0$, $\omega_1=\sqrt{\omega_0^2-\gamma^2}$, $\omega_2=\sqrt{3\omega_0^2-\gamma^2}$ and we take that the damping coefficient is in $\omega_0$ units, while the time is in $\omega_0^{-1}$ units. We are now in a position to determine the optimal damping of this 2-DOF system for different initial conditions. Again, we will not investigate all possible types of the initial conditions, but two qualitatively different ones, one with initial energy comprised only of potential energy, and the other with initial energy comprised only of kinetic energy. These two examples will give us a picture of the procedure for determining the optimal damping coefficient $\tilde{\gamma}$ for this 2-DOF system. The same procedure for determining the optimal damping can be in principle carried out for any MDOF system with MPD, with any initial condition.  

\subsubsection{Optimal damping of the 2-DOF system with initial energy comprised only of potential energy}

\begin{figure}[h!t!]
\begin{center}
\includegraphics[width=0.55\textwidth]{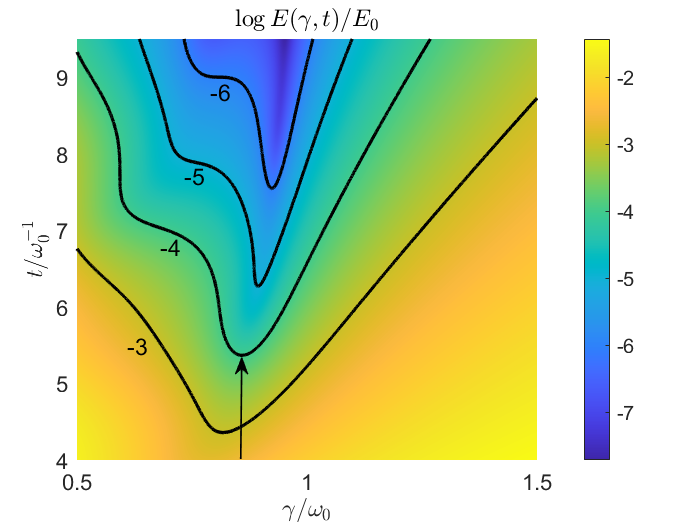}
\end{center}
\caption{The base $10$ logarithm of the ratio \eqref{Et2D}, i.e. $\log\left(E(\gamma, t)/E_{0}\right)$, for initial condition $E_{01}=E_{02}=E_0/2$ and $\theta_1=\theta_2=0$. For this initial condition, initial energy of the 2-DOF system is comprised only of potential energy distributed equally between the modes. Four black contour lines denote points with $E(\gamma,t)/E_{0}=\lbrace 10^{-3},10^{-4},10^{-5},10^{-6}\rbrace$ respectively, as indicated by the numbers placed to the left of each contour line. As an example of determining the optimal damping for which the system reaches the desired energy level the fastest, i.e. with respect to the condition \eqref{conditiontau}, we draw the arrow that points to the coordinates $(\gamma,t)=(0.859\omega_{0},5.37\omega_{0}^{-1})$ for which the energy of the system reaches the level $E(\gamma,t)/E_{0}=10^{-4}$ the fastest. Thus, $\tilde{\gamma}=0.859\omega_{0}$ is the optimal damping coefficient to reach this energy level the fastest. 
}
\label{fig:EPmin}
\end{figure}

 Here we choose initial condition with $E_{01}=E_{02}=E_0/2$ and $\theta_1=\theta_2=0$, i.e. with initial potential energy distributed equally between the two modes and zero initial kinetic energy. In Fig.\ \ref{fig:EPmin} we show the base $10$ logarithm of the ratio \eqref{Et2D}, i.e. $\log\left(E(\gamma, t)/E_0\right)$, for the chosen initial condition. 
 In Table \ref{table2D1} we show results for other energy levels corresponding to contour lines shown in Fig.\ \ref{fig:EPmin}. For this initial condition, optimal damping coefficient given by the minimization of the energy integral, i.e. \eqref{gammaND}, is $\gamma_{\opt}=\sqrt{3/4}\omega_{0}=0.866\omega_{0}$.
\begingroup
\setlength{\tabcolsep}{5.2pt} 
\renewcommand{\arraystretch}{1.5} 
\begin{table}[h!t!]
\centering
\begin{tabular}{|c|c|c|} 
 \hline
 $E(\gamma,t)/E_{0}$ & $\tilde{\gamma} \left[ \omega_{0} \right]$ & $\tau [ \omega_{0}^{-1}]$ \\ 
 \thickhline
 $10^{-3}$ & \ $0.817$ \ & $4.36$ \\ 
 \hline
 $10^{-4}$ & \ $0.859$ \ & $5.37$ \\ 
 \hline
 $10^{-5}$ & \ $0.893$ \ & $6.27$ \\ 
 \hline
 $10^{-6}$ & \ $0.924$ \ & $7.55$ \\ 
 \hline
\end{tabular}
\caption{Optimal damping coefficient $\tilde{\gamma}$ for which the energy of the system drops to the level $10^{-\delta}E_{0}$ the fastest, with the initial condition $E_{01}=E_{02}=E_0/2$ and $\theta_1=\theta_2=0$. 
}
\label{table2D1}
\end{table}
\endgroup

\subsubsection{Optimal damping of the 2-DOF system with initial energy comprised only of kinetic energy}

\begin{figure}[h!t!]
\begin{center}
\includegraphics[width=0.48\textwidth]{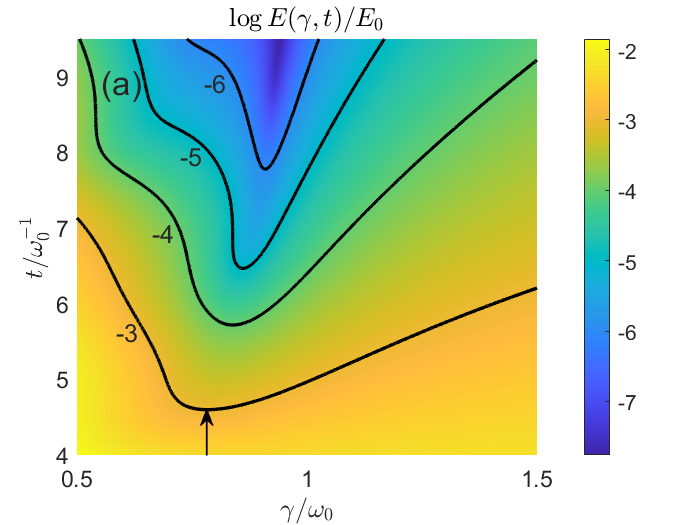}
\includegraphics[width=0.48\textwidth]{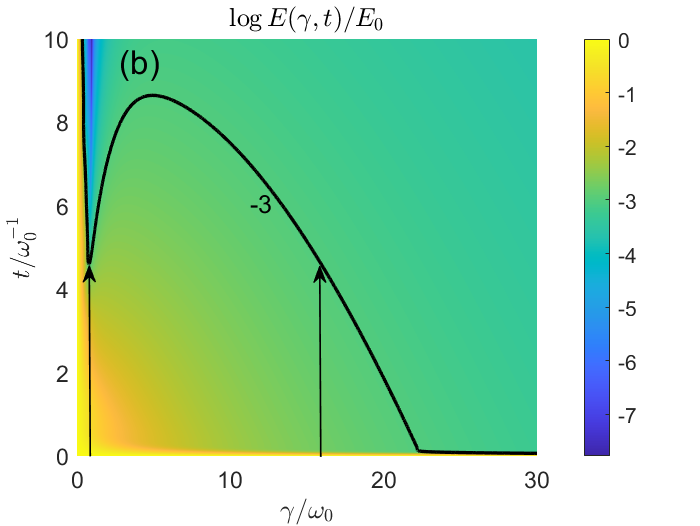}
\end{center}
\caption{The base $10$ logarithm of the ratio \eqref{Et2D}, i.e. $\log\left(E(\gamma, t)/E_{0}\right)$, for initial condition $E_{01}=E_{02}=E_0/2$ and $\theta_1=\theta_2=\pi/2$. For this initial condition, initial energy of the 2-DOF system is comprised only of kinetic energy distributed equally between the modes. (a) Four black contour lines denote points with $E(\gamma,t)/E_{0}=\lbrace10^{-3},10^{-4},10^{-5},10^{-6}\rbrace$ respectively, and the arrow points to the coordinates $(\gamma,t)=(0.783\omega_{0},4.60\omega_{0}^{-1})$, with $E(\gamma,t)/E_{0}=10^{-3}$, for which this level of energy is reached in shortest time for the shown data span. (b) Contour line for points with $E(\gamma, t)=10^{-3}E_{0}$ is shown for larger data span, left arrow points to the coordinates $(\gamma,t)=(0.783\omega_{0},4.60\omega_{0}^{-1})$, and the right arrow to the coordinates $(\gamma,t)=(15.927\omega_0,4.60\omega_0^{-1})$, both with $E_i(\gamma,t)/E_{0i}=10^{-3}$. Thus, for $\gamma>15.927\omega_{0}$ energy level $10^{-3}E_{0}$ is reached faster than for $\gamma=0.783\omega_{0}$.}
\label{fig:EKmin}
\end{figure}

 Here we choose initial condition with $E_{01}=E_{02}=E_0/2$ and $\theta_1=\theta_2=\pi/2$, i.e. with initial kinetic energy distributed equally between the two modes and zero initial potential energy. In Fig.\ \ref{fig:EKmin}(a) and (b) we show the base $10$ logarithm of the ratio \eqref{Et2D}, i.e. $\log\left(E(\gamma, t)/E_0\right)$, for the chosen initial condition. 
 In Table \ref{table2D2} we show results for other energy levels corresponding to contour lines shown in Fig.\ \ref{fig:EKmin}(a). For this initial condition, optimal damping coefficient given by the minimization of the energy integral, i.e. \eqref{gammaND}, is $\gamma_{\opt}=+\infty$.

\begingroup
\setlength{\tabcolsep}{5.2pt} 
\renewcommand{\arraystretch}{1.5} 
\begin{table}[h!t!]
\centering
\begin{tabular}{|c|c|c|} 
 \hline
 $E(\gamma,t)/E_{0}$ & $\tilde{\gamma} \left[ \omega_{0} \right]$ & $\tau [ \omega_{0}^{-1}]$ \\ 
 \thickhline
 $10^{-3}$ & \ $0.783$ \ & $4.60$ \\ 
 \hline
 $10^{-4}$ & \ $0.838$ \ & $5.72$ \\ 
 \hline
 $10^{-5}$ & \ $0.861$ \ & $6.47$ \\ 
 \hline
 $10^{-6}$ & \ $0.909$ \ & $7.78$ \\ 
 \hline
\end{tabular}
\caption{Optimal damping coefficient $\tilde{\gamma}$ for which the energy of the system drops to the level $10^{-\delta}E_{0}$ the fastest, with the initial condition $E_{01}=E_{02}=E_0/2$ and $\theta_1=\theta_2=\pi/2$. 
}
\label{table2D2}
\end{table}
\endgroup


\section{Conclusion and outlook}


The main message of this paper is that the dissipation of the initial energy in vibrating systems significantly depends on the initial conditions with which the dynamics of the system started, and ideally it would be optimal to always adjust the damping to the initial conditions. We took one of the known criteria for optimal damping, the criterion of minimizing the (zero to infinity) time integral of the energy of the system, averaged over all possible initial conditions corresponding to the same initial energy, and modified it to take into account the initial conditions, i.e. instead of averaging over of all possible initial conditions, we studied the dependence of the time integral of the energy of the system on initial conditions and determined the optimal damping as a function of the initial conditions. We found that the thus obtained optimal damping coefficients take on an infinite range of values depending on the distribution of initial potential energy and initial kinetic energy within the modes. We also pointed out the shortcomings of the thus obtained optimal damping coefficients and introduced a new method for determining optimal damping. Our method is based on the determination of the damping coefficients for which the energy of the system drops the fastest below some energy threshold (e.g. below the energy resolution of the experiment). We have shown that our method gives, both quantitatively and qualitatively, different results from the energy integral minimization method. In particular, the energy integral minimization method gives infinite optimal damping for initial conditions with purely kinetic energy, i.e. this method overlooks the region of underdamped coefficients for which strong energy dissipation occurs with this type of initial conditions, while this region is clearly seen and taken into account if one looks the energy behaviour directly, as we did. Furthermore, the energy integral minimization method gives the optimal damping which does not depend on the signs of the initial conditions, and we have shown that energy dissipation can strongly depend on them, which is taken into account in our method. 

Although the paper is dedicated to the case of mass-proportional damping, the new method we propose for determining the optimal damping can be applied to the types of damping we did not study in this paper. For example, in the case of a system with Rayleigh damping, the energy can be determined analytically using modal analysis, and based on that analytical expression, it can be numerically investigated for which values of the mass and stiffness proportionality constants the energy of the system drops the fastest below some energy threshold which effectively corresponds to the equilibrium state. In the case of a system with damping that does not allow analytical treatment, energy, as a function of time and magnitudes of individual dampers, can be determined numerically, e.g. by studying the vibrating system as a first order ordinary differential equation with matrix coefficients and using modern numerical methods for finding a solution of such an equation. This approach allows one to numerically solve systems with many degrees of freedom.   
Thus, we can numerically analyze the time evolution of the energy and find a set of damping parameters for which the energy drops to a desired energy threshold the fastest. Of course, this approach can be applied only for systems with a moderate number of degrees of freedom and a small number of dampers, due to 
the rapid growth of the parameter space that needs to be searched.
Despite these limitations, we believe that our approach to optimal damping can be useful because, as we have shown, it can provide insights that other approaches overlook. Therefore, in future work we will investigate in detail the application of our approach to systems with damping that does not allow modal analysis. Furthermore, real systems can respond to many different initial conditions in operating conditions. We envision that our approach can be used to provide an overall optimal damping with respect to all initial conditions or with respect to some expected range of initial conditions. For this purpose, one could consider the energy averaged over the initial conditions and find the damping for which this averaged energy drops to a desired energy threshold the fastest. This will be the topic of our next work. 

\section{Acknowledgments}

We are grateful to Bojan Lončar for making schematic figures of 2-DOF and MDOF systems, i.e. Fig.\ \ref{fig:skica2D} and \ref{fig:skicaND}, according to our sketches. This work was supported by the QuantiXLie Center of Excellence, a project co-financed by the Croatian Government and European Union through the European Regional Development Fund, the Competitiveness and Cohesion Operational Programme (Grant No. KK.01.1.1.01.0004).

The authors have no conflicts to disclose.

\appendix

\section{Average of the integral \eqref{intND2} over a set of all initial conditions}
\label{Appendix1}

For reader's convenience, we will repeat the integral \eqref{intND2} here
\begin{equation}
I(\gamma, \lbrace a_i\rbrace, \lbrace \theta_i\rbrace)=E_0\sum_{i=1}^N\frac{a_i^2}{2\omega_{0i}}\left(\frac{\omega_{0i}^2+\gamma^2}{\gamma\omega_{0i}}+\frac{\gamma}{\omega_{0i}}\cos2\theta_i+\sin2\theta_i\right)\,.
\label{intND2A}
\end{equation}
In order to calculate the average of \eqref{intND2A} over a set of all initial conditions, one has to integrate \eqref{intND2A} over all coefficients $a_i$, which satisfy $\sum_{i=1}^Na_i^2=1$ and $a_i\in[-1,1]$, and over all angles $\theta_i\in[0,2\pi]$. Due to $\int_0^{2\pi}\cos2\theta_id\theta_i=\int_0^{2\pi}\sin2\theta_id\theta_i=0$, terms with sine and cosine functions don't contribute to the average of \eqref{intND2A}. Integration over all possible coefficients $a_i$ amounts to calculating the average of $a_i^2$ over a sphere of radius one embedded in $N$ dimensional space. If we were to calculate the average of the equation of a sphere $\sum_{i=1}^Na_i^2=1$ over a sphere defined by that equation, we would get
\begin{equation}
\sum_{i=1}^N\overline{a_i^2}=1\,,
\label{prosjek}
\end{equation}
where $\overline{a_i^2}$ denotes the average of $a_i^2$ over a sphere. Due to the symmetry of the sphere and the fact that we are integrating over the whole sphere, contribution of each $\overline{a_i^2}$ in the sum \eqref{prosjek} has to be the same, so we can easily conclude that 
\begin{equation}
\overline{a_i^2}=\frac{1}{N}\,,
\label{prosjek1}
\end{equation}
for any $i$. Thus, the average of \eqref{intND2A} over all possible initial conditions is 
\begin{equation}
\overline{I}(\gamma)=\frac{E_0}{2N}\sum_{i=1}^N\left(\frac{\omega_{0i}^2+\gamma^2}{\gamma\omega_{0i}^2}\right)\,.
\label{intNDavA}
\end{equation}

\section{Limit values \eqref{limes1}, \eqref{limes2} and \eqref{limes3}}
\label{Appendix2}

For reader's convenience, we repeat here \eqref{avgammaNDopt} and \eqref{modoviN}
\begin{equation}
\overline{\gamma}_{\opt}=N^{1/2}\left(\sum_{i=1}^N\frac{1}{\omega_{0i}^2}\right)^{-1/2}\, 
\label{avgammaNDoptA}
\end{equation}
\begin{equation}
    \omega_{0i}=2\omega_0\sin\left(\frac{i\pi}{2(N+1)}\right)\,,\textrm{with}\,\,i=\lbrace1,...,N\rbrace\,.
    \label{modoviNA}
\end{equation}
Using \eqref{modoviNA}, we can write \eqref{avgammaNDoptA} as
\begin{equation}
\overline{\gamma}_{\opt}=2\omega_0 N^{1/2}\left(\sum_{i=1}^N\frac{1}{\sin^2\zeta_i}\right)^{-1/2}\,, 
\label{avgammaNDoptA1}
\end{equation}
 where $\zeta_i=\frac{i\pi}{2(N+1)}$. Using the fact that $\sin x < x$ for $0<x<\pi/2$, we obtain 
\begin{equation*}
    \overline{\gamma}_{\opt} < 2\omega_0 N^{1/2} \left( \frac{4 (N+1)^2}{\pi^2} \sum_{i=1}^N \frac{1}{i^2} \right)^{-1/2}
    = 2\omega_0 N^{1/2} \frac{\pi}{2(N+1)}\left(  \sum_{i=1}^N \frac{1}{i^2} \right)^{-1/2}.
\end{equation*}
 Now taking $N\to \infty$ and using the well-known formula $\sum_{i=1}^\infty \frac{1}{i^2}= \frac{\pi^2}{6}$, we obtain  \eqref{limes1}.  
%

Now we focus on the limit \eqref{limes2}. 
We will use the following well-known inequality $\sin x > x/2$ for $0<x<\pi/2$ (this can be easily seen by, e.g.\ using the fact that $\sin$ is a concave function on $[0,\pi/2]$). From \eqref{avgammaNDoptA1} it follows
\begin{equation*}
    \frac{\overline{\gamma}_{\opt}}{\omega_{01}} = N^{1/2} (\sin \zeta_1)^{-1} \left(\sum_{i=1}^N\frac{1}{\sin^2\zeta_i}\right)^{-1/2} >
    N^{1/2} \zeta_1^{-1} \cdot \frac{1}{2}   \left(\sum_{i=1}^N\frac{1}{\zeta_i^2}\right)^{-1/2} = \frac{1}{2} N^{1/2} \left(\sum_{i=1}^N\frac{1}{i^2}\right)^{-1/2},
\end{equation*}
hence we obtain \eqref{limes2}. 

The limit \eqref{limes3} is also easy to prove. Since
\begin{equation}
\lim_{N \to +\infty}\omega_{0N}=\lim_{N \to +\infty}2\omega_0\sin\left(\frac{N\pi}{2(N+1)}\right)=2\omega_0\,
\label{dostamije}
\end{equation}
and we already showed \eqref{limes1}, it is easy to conclude that
\begin{equation}
\lim_{N \to +\infty}\frac{\omega_{0N}}{\overline{\gamma}_{\opt}}=+\infty\,, 
\end{equation}
i.e. the limit \eqref{limes3} holds.

\bibliographystyle{elsarticle-num}  
\bibliography{aipsamp}

\end{document}